\documentclass[pra,twocolumn,aps,amssymb,showpacs,superscriptaddress]{revtex4-1}

\usepackage{epsfig}
\usepackage{graphicx}
\usepackage{amsmath}
\usepackage{amssymb}
\usepackage{enumerate}

\usepackage{color}
\definecolor{rosso}{rgb}{1,0,0}
\definecolor{verde}{rgb}{0,1,0}
\definecolor{blue}{rgb}{0,0,1}
\definecolor{verdescuro}{rgb}{0,0.5,0.5}
\definecolor{rossoscuro}{rgb}{0.7,0.3,0}
\definecolor{bluscuro}{rgb}{0.3,0,0.7}
\definecolor{magenta}{rgb}{1,0,1}

\begin{document}

\title{Entanglement between pairing and screening in the Gorkov-Melik-Barkhudarov correction to the critical temperature throughout the BCS-BEC crossover}

\author{L. Pisani}
\affiliation{School of Science and Technology, Physics Division, Universit\`{a} di Camerino, 62032 Camerino (MC), Italy}
\author{A. Perali}
\affiliation{School of Pharmacy, Physics Unit, Universit\`{a} di Camerino, 62032 Camerino (MC), Italy}
\author{P. Pieri}
\email{pierbiagio.pieri@unicam.it}
\affiliation{School of Science and Technology, Physics Division, Universit\`{a} di Camerino, 62032 Camerino (MC), Italy}
\affiliation{INFN, Sezione di Perugia, 06123 Perugia (PG), Italy}
\author{G. Calvanese Strinati}
\email{giancarlo.strinati@unicam.it}
\affiliation{School of Science and Technology, Physics Division, Universit\`{a} di Camerino, 62032 Camerino (MC), Italy}
\affiliation{INFN, Sezione di Perugia, 06123 Perugia (PG), Italy}

\date{\today}

\begin{abstract}
The problem of the theoretical description of the critical temperature $T_{c}$ of a Fermi superfluid dates back to the work by Gor'kov and Melik-Barkhudarov (GMB), who addressed it for a weakly-coupled (dilute) superfluid in what would today be referred to as the (extreme) BCS (weak-coupling) limit of the BCS-BEC crossover.
The point made in this context by GMB was that particle-particle (pairing) excitations, which are responsible for superfluidity to occur below $T_{c}$, and particle-hole excitations, which give rise to screening also in a normal system, get effectively disentangled from each other in the BCS limit, thus yielding a reduction by a factor $2.2$ of the value of $T_{c}$ obtained when neglecting screening effects.
Subsequent work on this topic, that was aimed at extending the original GMB argument away from the BCS limit with diagrammatic methods, has \emph{tout court} kept this disentangling between pairing and screening throughout the BCS-BEC crossover, without realising that the conditions for it to be valid are soon violated away from the BCS limit.
Here, we reconsider this problem from a more general perspective and argue that pairing and screening are intrinsically entangled with each other along the whole BCS-BEC crossover but for the BCS limit considered by GMB, with the particle-hole excitations soon transmuting into particle-particle excitations away from this limit.
We substantiate our argument by performing a detailed numerical calculation of the GMB diagrammatic contribution suitably extended to the whole BCS-BEC crossover, where the full wave-vector and frequency dependence occurring in the repeated in-medium two-particle scattering is duly taken into account for the first time.
Our numerical calculations are tested against analytic results available in both the BCS and BEC limits, and the contribution of the GMB diagrammatic term to the scattering length of composite bosons in the BEC limit is highlighted.
We calculate $T_{c}$ throughout the BCS-BEC crossover and find that it agrees quite well with Quantum Monte Carlo calculations and experimental data available in the unitarity regime.
\end{abstract}

                 
\maketitle

\section{Introduction} 
\label{sec:introduction}

In the original work by Bardeen, Cooper, and Schrieffer (BCS) on the theory of superconductivity \cite{BCS-1957}, the attractive inter-particle interaction of strength $U_{0}$ acting between opposite-spin fermions was considered to affect an energy region about the Fermi surface with width of the order of the Debye frequency $\omega_{D}$ (we set $\hbar = 1$ throughout).
This led to a critical temperature for the onset of superconductivity, of the form 
\begin{equation}
k_{B} T_{c}^{\mathrm{BCS}} = \frac{2 e^{\gamma} \omega_{D}}{\pi} \exp\{-1/(N_{0}|U_{0}|)\}
\label{Tc-BCS-cutoff}
\end{equation}
\noindent 
where $N_{0}= m k_{F}/(2 \pi^{2})$ is the density of states at the Fermi level per spin component ($m$ being the fermion mass and $k_{F} = \left(3 \pi^{2} n \right)^{1/3}$ the Fermi wave vector associated with the particle density $n$), $e$ the Euler number, and $k_{B}$ and $\gamma$ the Boltzmann and Euler constants (with $e^{\gamma} \simeq 1.781$).

Soon after the BCS paper, Gork'ov and Melik-Barkhudarov (GMB) considered the phenomenon of superfluidity in a dilute (neutral) Fermi gas \cite{GMB-1961}.
This physical system has the advantage over the BCS model, that the effects of the inter-particle interaction can be expressed entirely in terms of the scattering length $a_{F}$ of the two-fermion problem in vacuum, thereby leaving aside the uncertainties related to the strength $U_{0}$ and the cutoff $\omega_{D}$.
The dilute Fermi gas was originally considered by Galitskii for the case of a repulsive inter-particle interaction, for which $a_{F} > 0$ and $k_{F} a_{F} \ll 1$ \cite{Galitskii-1958}.
To deal with the phenomenon of superfluidity, GMB extended this treatment to the case of an attractive inter-particle interaction, for which $a_{F} < 0$ and $k_{F} |a_{F}| \ll 1$.
Nowadays, full experimental control of the fermionic scattering length $a_{F}$ can be achieved with ultra-cold Fermi gases with an attractive inter-particle interaction \cite{Regal-2003}.

Having disposed off the quantities $U_{0}$ and $\omega_{D}$ with their associated uncertainties, the critical temperature for a dilute Fermi gas can still be obtained within the BCS mean-field decoupling, leading to the expression
\begin{equation}
k_{B} T_{c}^{\mathrm{BCS}} = \frac{8 e^{\gamma} E_{F}}{\pi e^{2}} \exp\{\pi/(2 k_{F} a_{F})\}
\label{Tc-BCS-no_cutoff}
\end{equation} 
where $E_{F} = k_{F}^{2}/(2m)$ is the Fermi energy.
[Note that the expression (\ref{Tc-BCS-no_cutoff}) can be formally obtained from the original BCS expression (\ref{Tc-BCS-cutoff}) with the replacements $\omega_{D} \rightarrow 4 E_{F}/e^{2}$ and  $U_{0} \rightarrow 4 \pi a_{F}/m$.]
What GMB then realized was that, owing to the exponential dependence of the BCS expression (\ref{Tc-BCS-no_cutoff}) for $T_{c}$, if additional terms in the small parameter $k_{F} a_{F}$ could be introduced in the exponent such that $(k_{F} a_{F})^{-1} \rightarrow (k_{F} a_{F})^{-1} + b + c \, (k_{F} a_{F}) + \cdots$ where $b$ and $c$ are constants, the constant $b$ would modify the BCS pre-factor of $T_{c}$ by a finite amount even in the (extreme) weak-coupling limit when $k_{F} a_{F} \rightarrow 0^{-}$.
To obtain the constant $b$, GMB considered a correction to the BCS instability of the normal phase that occurs when $T_{c}$ is approached from above.
This instability can be obtained diagrammatically in terms of ``the series of ladder graphs'' in the particle-particle channel \cite{Schrieffer-1964}, and yields correspondingly for $T_{c}$ the BCS mean-field result (\ref{Tc-BCS-no_cutoff}) as obtained when $T_{c}$ is approached from below. 
For this reason, the GMB correction is sometimes referred to as ``beyond-mean-field approximation''.
The end result of the GMB calculation for $T_{c}$ was a reduction of the expression (\ref{Tc-BCS-no_cutoff}) of $T_{c}$ by the factor $(4e)^{1/3} \simeq 2.2$.
This result was obtained by performing, in practice, all wave-vector integrations contained in the diagrammatic expressions near the Fermi surface, with the assumption that the fermionic chemical potential $\mu$ coincides with $E_{F}$.

Later on, the GMB effect on $T_{c}$ was interpreted on physical grounds in terms of ``polarization'' effects of the medium occurring in the particle-hole channel due to particle-hole excitations across the Fermi surface \cite{Heiselberg-2000}, and attempts along these lines were also made to extend the GMB result to the high-density regime \cite{Combescot-1999}.
In all cases, these calculations were still limited to the (weak-coupling) BCS limit where the non-interacting Fermi surface is only slightly perturbed by the inter-particle attraction.
In this limit (as we shall comment more extensively below), the primary BCS instability occurring in the particle-particle channel and the polarization effects of the medium occurring in the particle-hole channel get effectively disentangled from each other (with an accompanying large reduction in the computational effort for determining their effect on $T_{c}$).
Extensions of the GMB result were also considered for lower dimensionality \cite{Petrov-2003},
for mixtures of two-component fermionic atoms with different masses \cite{Baranov-2008},
for imbalanced spin populations in quasi-two dimensions \cite{Resende-2012}, 
and for lattice models \cite{Kim-2009},
but in all cases still adopting the standard approximations that apply to the BCS regime.

With the advent of the experiments with ultra-cold Fermi gases and the related study of the BCS-BEC crossover, however, the question has naturally arisen about what would be the effect of the GMB correction when departing from the BCS limit. 
In particular, it is of interest to assess whether this correction may still yield significant effects at the unitary limit $(k_{F} a_{F})^{-1} = 0$ where a remnant Fermi surface is still active \cite{Bulgac-2006}.
By the BCS-BEC crossover, there occurs a progressive reduction of the size of the fermionic pairs, ranging from the large size of strongly overlapping Cooper pairs in the BCS limit of weak inter-particle attraction, to the small size of non-overlapping composite bosons in the BEC limit of a strong inter-particle attraction, across the intermediate unitary limit where the size of the pairs is comparable with the average inter-particle distance.
This crossover is driven by the coupling parameter $(k_F a_F)^{-1}$, which ranges from  $(k_F\, a_F)^{-1} \lesssim -1$ in the weak-coupling (BCS) regime when $a_F < 0$, to $(k_F\, a_F)^{-1} \gtrsim +1$ 
in the strong-coupling (BEC) regime when $a_F > 0$, across the unitary limit when $|a_F|$ diverges.    
Correspondingly, the fermionic chemical potential $\mu$ ranges from $E_{F}$ in the BCS limit to $-(2 m a_{F}^{2})^{-1}$ in the BEC limit, with a large variation occurring in between these two limits.

In this context, when dealing diagrammatically with the GMB correction to $T_{c}$ throughout the BCS-BEC crossover, the effective disentangling between particle-particle and particle-hole channels mentioned above, which should confidently apply to the BCS limit only, was instead carried over \emph{tout court} to the whole crossover \cite{Yu-2009,Ruan-2013}.
In practice, this was simply done by calculating the particle-hole bubble associated with particle-hole excitations (suitably averaged over the Fermi sphere, like in the original GMB calculation), but now with a chemical potential that spans the whole crossover and thus is no longer equal to $E_{F}$ (apart from minor differences resulting from the way the chemical potential itself is calculated \cite{Yu-2009,Ruan-2013}).
Recently, extensions along these lines were also considered to investigate the GMB correction when including the effect of the Rashba spin-orbit coupling in two-dimensional Fermi gases \cite{Lee-2017}. 
A completely different approach was instead followed in Ref.~\cite{Floerchinger-2008}, where particle-particle and particle-hole bubbles were included simultaneously in the framework 
of the functional-renormalization-group approach.

Purpose of this paper is to analyze and settle the question of entanglement vs disentanglement between pairing and screening in the GMB correction to the critical temperature throughout the \emph{whole} 
BCS-BEC crossover, by a careful analysis of the relevant many-body diagrammatic structure of the theory.
Our analysis combines both extensive numerical calculations applied to the whole crossover and analytic results in the BCS and BEC limits, and avoids at the outset the approximations introduced originally by GMB. 
We will show that these approximations, which entail an effective disentanglement between pairing (in the particle-particle channel) and screening (in the particle-hole channel), hold only in the BCS limit where 
the particle-particle propagator is approximately constant over a large sector of the wave-vector and frequency domain.
Away from the BCS limit, the particle-particle (pair) propagator acquires instead a progressively \emph{marked dependence on wave vector and frequency\/} and the GMB disentanglement between pairing and screening no longer holds.
This feature makes the numerical calculation of the GMB correction to the critical temperature quite more involved than those reported, e.~g., in Refs.~\cite{Yu-2009,Ruan-2013} where the GMB disentanglement was assumed to hold for the whole crossover.
In addition, for a realistic calculation of the critical temperature throughout the whole BCS-BEC crossover, we have combined the GMB correction, which evolves from the BCS to the BEC limits, with another correction which instead evolves in the opposite direction from the BEC to the BCS limits, since it was conceived in Ref.~\cite{Pieri-2005} to improve on the description of composite bosons in the BEC limit, at the level of the Popov approximation for point-like bosons \cite{Popov-1987}.
This combined Popov-GMB calculation for the critical temperature throughout the whole BCS-BEC crossover will yield a quite good comparison with Quantum Monte Carlo data over the whole coupling range for which they are available and with experimental data at unitarity.

Finally, it is relevant to mention that interest in the effects of medium polarization has also arisen in the context of low-density neutron matter, to the extent that these effects affect the value of the pairing gap in the superfluid phase at zero temperature \cite{Schulze-2001,Cao-2006}.
Actually, it was already shown in the original GMB paper \cite{GMB-1961} that their beyond-mean-field approximation also renormalizes the value of the pairing gap at zero temperature with respect to the BCS value, in quite the same way that it does for the value of the critical temperature.
The question of the entanglement between pairing and screening away from the weak-coupling (BCS) limit, that we discuss in detail in this paper for the critical temperature, is then expected to be relevant for the pairing gap as well and will be considered in future work.

The plan of the paper is as follows.
In Section~\ref{sec:G-MB-BCS-BEC} we set up a diagrammatic approach to pairing fluctuations above $T_{c}$, which includes the GMB contribution in a form that can be extended to the whole BCS-BEC crossover.
In this way, we show that in the BCS limit pairing and screening get effectively disentangled from each other, thus recovering the original GMB result.
We also show, however, that this disentanglement cannot be sustained away from the BCS limit. 
To this end, we calculate analytically how the GMB correction evolves toward the BEC limit, showing that in this limit it yields a significant contribution to the scattering length for composite bosons. 
We further discuss the need to introduce the Popov correction mentioned above.
In Section~\ref{sec:numerical-results} we report on the numerical calculation of the Popov and GMB corrections and show how the relevant quantities in this context behave across the BCS-BEC crossover. 
We also discuss a (partial) self-consistent procedure on the pair propagator which is required for the calculation of $T_{c}$, owing to the fact that this propagator diverges upon approaching the superfluid phase when lowering the temperature from the normal phase.
The numerical results for the Popov and GMB corrections are tested against their analytic expressions of Section~\ref{sec:G-MB-BCS-BEC} in the BCS and BEC limits, and our results for $T_{c}$ throughout the BCS-BEC crossover are compared with the Quantum Monte Carlo and experimental data available in the unitary regime.
Section~\ref{sec:conclusions} gives our conclusions and sets up future perspectives of our approach.
Appendix~\ref{sec:appendix-A} discusses in detail the way the numerical calculations of the Popov and GMB corrections have been implemented in practice, in view of the highly non-trivial task of including the full wave-vector and frequency dependence of the pair propagators that enter these corrections.
Finally, Appendix~\ref{sec:appendix-B} shows how the Popov and GMB corrections contribute to the value of the scattering length for composite bosons that form in the BEC limit.

\section{Extending the GMB contribution throughout the BCS-BEC crossover} 
\label{sec:G-MB-BCS-BEC}

In this Section, we show how the GMB contribution has to be handled away from the BCS (weak-coupling) limit of the BCS-BEC crossover, in the normal phase above the critical temperature $T_{c}$.
To this end, we begin by discussing the essential aspects of the many-body diagrammatic theory for a dilute Fermi gas with an attractive inter-particle interaction, which are relevant to this problem.
We shall explicitly be concerned about determining the value of $T_{c}$ throughout the BCS-BEC crossover, by resting on \emph{a minimal set} of diagrammatic terms which include the GMB contribution.
In the following, both the reduced Planck constant $\hbar$ and the Boltzmann constant $k_{B}$ are set equal to unity.

\vspace{0.05cm}
\begin{center}
{\bf A. Brief summary about pairing fluctuations \\ in the normal phase above $T_{c}$}
\end{center}
\vspace{-0.2cm}

A dilute Fermi gas is characterised by the fact that the range of the inter-particle interaction is much smaller than the average inter-particle distance, such that the interaction can be taken
of the contact type $v_{0} \delta(\mathbf{r}-\mathbf{r'})$ acting between opposite-spin fermions.
In the following, the attractive case $v_{0}<0$ will only be considered and equal spin populations will be taken.

The choice of a contact potential entails to introduce an ultraviolet cutoff $k_{0}$ in the otherwise divergent integrals over the wave vector $\mathbf{k}$. 
The two quantities $v_{0}$ and $k_{0}$ can be combined together by resorting to the two-body problem in vacuum, whereby the fermionic scattering length $a_{F}$ is obtained from the relation
\cite{Sa_de_Melo-1993}:
\begin{equation}
\frac{m}{4\pi a_{F}} =  \frac{1}{v_{0}} + \int_{|\mathbf{k}| \le k_{0}} \! \frac{d\mathbf{k}}{(2 \pi)^3} \, \frac{m}{\mathbf{k}^2} \, \, .
\label{regularization}
\end{equation}
\noindent
A suitable \emph{regularization procedure} can be introduced at this point which eliminates further reference to $v_{0}$ and $k_{0}$, by taking the limits $v_{0} \rightarrow 0^{-}$ and 
$k_{0} \rightarrow \infty$ simultaneously such that $a_{F}$ remains fixed at a desired value.

\begin{figure}[t]
\begin{center}
\includegraphics[width=7.5cm,angle=0]{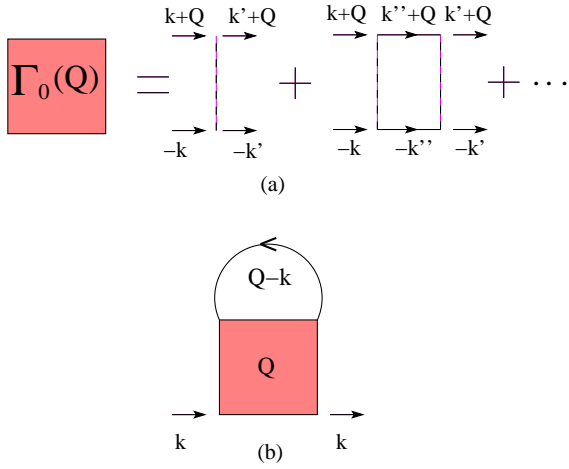}
\caption{(Color online) Diagrammatic representation of: 
                                    (a) The bare pair propagator $\Gamma_{0}$ for opposite-spin fermions; 
                                    (b) The single-particle fermionic self-energy $\Sigma$ obtained from $\Gamma_{0}$ within the (non-self-consistent) t-matrix approximation.
                                    Full and dashed lines represent the bare fermionic single-particle propagator $G_{0}$ and the interaction potential $v_{0}$, respectively, while $Q$ ($k$) 
                                    are bosonic (fermionic) four-vectors.
                                    Here and in the following, the upper (lower) line of $\Gamma_{0}$ corresponds to an up (down) fermionic spin.}
\label{Figure-1}
\end{center} 
\end{figure}

The above procedure is especially relevant when dealing with ultra-cold Fermi gases, for which $a_{F}$ can be experimentally controlled \cite{Regal-2003}.
From a theoretical side, this regularization procedure somewhat simplifies the structure of the many-body diagrammatic theory.
This is because a given diagram, which can be drawn for finite $v_{0}$, survives the limit $v_{0} \rightarrow 0^{-}$ \emph{provided\/} there occurs a compensating ultraviolet divergence for $k_{0} \rightarrow \infty$. 
It turns then out that in this way the bare interaction $v_{0}$ is everywhere replaced by the effective interaction (or bare \emph{pair propagator}) $\Gamma_{0}(Q)$ for opposite-spin fermions, where $Q=(\mathbf{Q},\Omega_{\nu})$ is a four-vector with bosonic Matsubara frequency $\Omega_{\nu}=2 \pi \nu T$ ($\nu$ integer).
The pair propagator, depicted in Fig.~\ref{Figure-1}(a), corresponds to an infinite sequence of two-body scattering events in the particle-particle channel (or ladder diagrams) and is given by the expression:
\begin{equation}
\Gamma_{0}(Q) = \frac{ - v_{0}}{1+v_{0} \, \chi_{\mathrm{pp}}(Q)} = - \frac{1}{\frac{m}{4 \pi a_F} + R_{\mathrm{pp}}(Q)} \,\, .
\label{Gamma-0}
\end{equation}
\noindent
Here, 
\begin{equation}
\chi_{\mathrm{pp}}(Q) = \int \! \frac{d\mathbf{k}}{(2\pi)^3} \, T \, \sum_{n} \, G_{0}(\mathbf{k}+\mathbf{Q},\omega_{n}+\Omega_{\nu}) \, G_{0}(-\mathbf{k},-\omega_{n})
\label{particle-particle-bubble}
\end{equation}
is the particle-particle bubble, where $G_{0}(\mathbf{k},\omega_{n}) = (i\omega_{n} - \xi_{\mathbf{k}})^{-1}$ is the bare fermionic single-particle propagator with $\xi_{\mathbf{k}}=\mathbf{k}^2/(2m) - \mu$ ($\mu$ being the chemical potential) and fermionic Matsubara frequency $\omega_{n}=(2 n + 1) \pi T$ ($n$ integer), and
\begin{eqnarray}
R_{\mathrm{pp}}(Q) & = & \chi_{\mathrm{pp}}(Q) - \int \!\frac{d\mathbf{k}}{(2 \pi)^3} \frac{m}{\mathbf{k}^2} 
\nonumber \\
& = & \int \!\frac{d\mathbf{k}}{(2 \pi)^3} \left( \frac{1 - f(\xi_{\mathbf{k}+\mathbf{Q}}) - f(\xi_{\mathbf{k}})} {\xi_{\mathbf{k}+\mathbf{Q}}+\xi_{\mathbf{k}}-i\Omega_{\nu}} - \frac{m}{\mathbf{k}^2} \right)
\label{bubble-pp-exact}  
\end{eqnarray}
\noindent
is the regularised version of the particle-particle bubble obtained with the help of Eq.~(\ref{regularization}) ($f(E) = (\exp\{E/T\} +1)^{-1}$ being the Fermi function).

Accordingly, the pair propagator $\Gamma_{0}$ represents a building block of the diagrammatic theory in the normal phase above $T_{c}$.
In particular, the simplest possible fermionic self-energy is obtained in terms of a single $\Gamma_{0}$, as depicted in Fig.~\ref{Figure-1}(b) and given by the expression:
\begin{equation}
\Sigma(k) = - \int \! \frac{d\mathbf{Q}}{(2 \pi)^3} \, T \sum_{\nu} \, \Gamma_{0}(Q) \, G_{0}(Q-k)
\label{Sigma-NSR} 
\end{equation}
\noindent
with the fermionic four-vector notation $k=(\mathbf{k},\omega_{n})$.
The choice (\ref{Sigma-NSR}) for $\Sigma$ corresponds to the so-called ``non-self-consistent t-matrix approximation''.
This self-energy was originally considered by Nozi\`{e}res and Schmitt-Rink (NSR), to account for the correct behavior of the BCS-BEC crossover in the BEC limit at finite temperature \cite{NSR-1985}.
In the following, we adopt the approach of Ref.\cite{PPSC-2002} and use the self-energy (\ref{Sigma-NSR}) to obtain the (dressed) fermionic single-particle propagator $G(k) = [G_{0}(k)^{-1} - \Sigma(k)]^{-1}$, in terms of which the total fermionic density \begin{equation}
n =  2  \int \! \frac{d\mathbf{k}}{(2 \pi)^3} \, T \, \sum_{n} \, e^{i\omega_{n}\eta} \, G(k)
\label{density}
\end{equation}
\noindent
(where $\eta = 0^{+}$) can be calculated to obtain the chemical potential for given temperature and coupling.
In the original NSR approach \cite{NSR-1985}, however, the density was derived from  Eq.~(\ref{density}) with the additional approximation of expressing $G$ at first order in $\Sigma$, thereby writing $G(k) \simeq G_{0}(k) + G_{0}(k) \, \Sigma(k) \, G_{0}(k)$.

\vspace{0.05cm}
\begin{center}
{\bf B. Thouless criterion}
\end{center}
\vspace{-0.2cm}

The pair propagator $\Gamma_{0}(Q)$ plays also the role of signaling the insurgence of the broken-symmetry (superfluid) phase, when the temperature is lowered down to the critical temperature $T_{c}$. 
This is because the sequence of ladder diagrams on which the pair propagator is build diverges for $Q=0$, thereby manifesting that upon approaching $T_{c}$ pairing fluctuations are able to organize themselves over a progressively larger spatial distance. 
The critical temperature is thus determined by the following condition (known also as Thouless criterion \cite{Thouless-1960}):
\begin{eqnarray}
& - & \Gamma_{0}(Q=0;T_{c},\mu_{c})^{-1} = \frac{m}{4 \pi a_F} 
\nonumber \\
& + & \int \! \frac{d\mathbf{k}}{(2 \pi)^3} \! \left( \! \frac{\tanh(\xi_{\mathbf{k}}/2T_{c}) }{2\xi_{\mathbf{k}}} - \frac{m}{\mathbf{k}^2} \! \right) = 0
\label{Thouless-criterion}
\end{eqnarray}
\noindent
obtained upon setting $Q=0$ in the expression (\ref{Gamma-0}) and with the thermodynamic variables explicitly indicated.
This expression formally coincides with that obtained within the BCS mean-field approximation when $T_{c}$ is reached from below.
Care should, however, be exerted about the value of $\mu_{c} = \mu(T_{c})$ that enters Eq.~(\ref{Thouless-criterion}).

For later purposes, it is relevant to obtain analytically the expression of $T_{c}$ in the BCS (weak-coupling) limit within the non-self-consistent t-matrix approximation of Eqs.~(\ref{Sigma-NSR}) and (\ref{density}).
To this end, the integral on the right-hand side of Eq.~(\ref{Thouless-criterion}) can be calculated analytically under the typical weak-coupling approximation $T_{c} \ll \mu$.
One obtains:
\begin{small}
\begin{equation}
\int \!\! \frac{d\mathbf{k}}{(2 \pi)^3} \! \left( \! \frac{\tanh(\xi_{\mathbf{k}}/2T_{c}) }{2\xi_{\mathbf{k}}} - \frac{m}{\mathbf{k}^2} \! \right) 
\simeq \! \frac{(2 m)^{3/2} \! \sqrt{\mu}}{4 \pi^{2}} \! \left[ \ln \left( \frac{8 \mu e^{\gamma}}{\pi T_{c}} \right) \! - 2 \right] \! .
\label{approximate-BCS-integral}
\end{equation}
\end{small}

\noindent
Entering this result into the Thouless criterion (\ref{Thouless-criterion}) yields:
\begin{equation}
T_{c} \simeq \frac {8 e^{\gamma} \mu}{\pi e^{2}} \exp \left\{ \frac{\pi}{2 k_{F} a_{F}} \sqrt{\frac{E_{F}}{\mu}} \right\} \, .
\label{Tc-non-self-consistent-t-matrix}
\end{equation}
\noindent
Within the non-self-consistent t-matrix approximation (as well as in the NSR approach), to the leading order in the small parameter $k_{F} |a_{F}|$ (where $a_{F} < 0$) the chemical potential 
(for $T \ll T_{F}$, as is the case in weak coupling close to $T_{c}$) is given by
\begin{equation}
\mu = E_{F} \left[ 1 + \frac{4}{3 \pi} k_{F} a_{F} + \cdots \right] \, ,
\label{mu-BCS-limit}
\end{equation}
\noindent
such that $\pi \sqrt{E_{F}/\mu} / (2 k_{F} a_{F}) \simeq \pi / (2 k_{F} a_{F}) - 1/3$ in the exponent of Eq.~(\ref{Tc-non-self-consistent-t-matrix}).
This results in a ``spurious'' factor $e^{-1/3}$ to appear on the right-hand side of Eq.~(\ref{Tc-non-self-consistent-t-matrix}) with respect to the expected BCS (weak-coupling) result (\ref{Tc-BCS-no_cutoff}).
This shortcoming can be remedied by introducing a partial degree of self-consistency in the non-self-consistent t-matrix approximation through a constant self-energy (mean-field) shift 
$\Sigma_{0} \simeq 2 \pi a_{F} n / m$ \cite{PPSC-2002}, such that $\mu - \Sigma_{0} \simeq E_{F}$ again at the leading order in $k_{F} |a_{F}|$.
In this way, the expression (\ref{Tc-BCS-no_cutoff}) for $T_{c}$ is correctly recovered.

In the following, a partial degree of self-consistency will be introduced in the non-self-consistent t-matrix approximation, not only in the BCS (weak-coupling) limit but also throughout 
the whole BCS-BEC crossover, by relying on the approach of Ref.\cite{Pieri-2005}.
This approach (to be discussed next) was originally conceived to improve on the description of composite bosons (dimers) that form in the BEC limit with respect to the non-self-consistent t-matrix approximation, by introducing a kind of mean-field interaction also between the otherwise non-interacting composite bosons.

\vspace{0.05cm}
\begin{center}
{\bf C. Popov contribution}
\end{center}
\vspace{-0.2cm}

\begin{figure}[t]
\begin{center}
\includegraphics[width=6.5cm,angle=0]{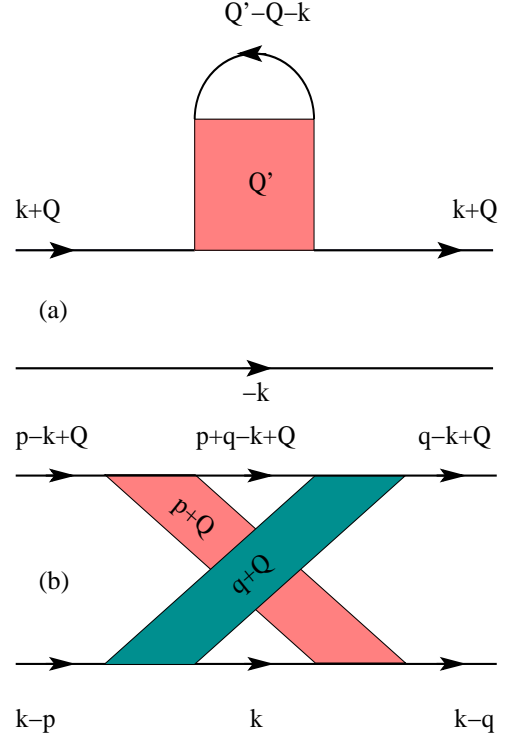}
\caption{(Color online)  Diagrammatic representation of: 
                                     (a) The Popov bosonic-like self-energy $\Sigma_{\mathrm{Popov}}^{\mathrm{B}}$, obtained by dressing the upper fermionic line in the particle-particle channel 
                                     (an analogous dressing occurs for the lower fermionic line);
                                     (b) The GMB bosonic-like self-energy $\Sigma_{\mathrm{GMB}}^{\mathrm{B}}$.
                                     In both cases, upper and lower fermionic lines correspond to opposite spins.
                                     By the present approach, $\Sigma_{\mathrm{Popov}}^{\mathrm{B}}$ and $\Sigma_{\mathrm{GMB}}^{\mathrm{B}}$ represent bosonic-like 
                                     self-energy insertions to the pair propagator $\Gamma_{0}$.}
\label{Figure-2}
\end{center} 
\end{figure}

In Ref.\cite{Pieri-2005}, an approximation for the BCS-BEC crossover was devised, which in the strong-coupling limit of the fermionic attraction would extend the Popov description from point-like \cite{Popov-1987}
to composite bosons.
In this way, a residual interaction among composite bosons survives even when the condensate disappears above the critical temperature.
Under these circumstances, the bare pair propagator $\Gamma_{0}$ gets dressed through the bosonic-like self-energy depicted in Fig.~\ref{Figure-2}(a), whose analytic expression reads \cite{footnote-1}:
\begin{eqnarray}
& & \Sigma_{\mathrm{Popov}}^{\mathrm{B}}(Q) = - 2  \int \! \frac{d\mathbf{k}}{(2 \pi)^3} \, T \, \sum_{n} \int \! \frac{d\mathbf{Q'}}{(2 \pi)^3} \, T \sum_{\nu'} 
\nonumber \\
& \times & G_{0}(k+Q)^{2} G_{0}(-k) G_{0}(Q'-Q-k) \Gamma_{0}(Q')
\label{Popov-self-energy-definition}
\end{eqnarray}
\noindent
where the factor of $2$ accounts for the dressing of both upper and lower fermionic lines.
In the following, we shall limit ourselves to consider the case $Q=0$, whereby we set $\Sigma_{\mathrm{Popov}}^{\mathrm{B}}=\Sigma_{\mathrm{Popov}}^{\mathrm{B}}(Q=0)$.

Both the BCS (weak-coupling) and BEC (strong-coupling) limits of $\Sigma_{\mathrm{Popov}}^{\mathrm{B}}$ can be calculated analytically.
In particular, the BCS limit is obtained as follows.
One begins by approximating $\Gamma_{0}(Q') \simeq - 4 \pi a_{F}/m$ in Eq.~(\ref{Popov-self-energy-definition}), such that
\begin{equation}
\int \! \frac{d\mathbf{Q'}}{(2 \pi)^3} \, T \sum_{\nu'} G_{0}(Q'-k) \Gamma_{0}(Q') \simeq - \frac{4 \pi a_{F}}{m} \, \frac{n_{0}(T,\mu)}{2}
\label{Popov-first-approximation} 
\end{equation}
\noindent
where $n_{0}(T,\mu)$ is the density of a system of non-interacting fermions with Fermi energy $\mu$ at temperature $T$.
Accordingly, for $T \ll T_{F}$ we can write $n_{0}(T,\mu) = k_{\mu}^{3}/(3 \pi^{2})$, with the notation $\mu = k_{\mu}^{2}/(2m)$ (with $\mu>0$) \cite{PPSC-2002}.
The remaining factor on the right-hand side of Eq.~(\ref{Popov-self-energy-definition}) can be calculated by noting that
\begin{small}
\begin{eqnarray}
& & \int \! \frac{d\mathbf{k}}{(2 \pi)^3} \, T \, \sum_{n} G_{0}(k)^{2} G_{0}(-k) 
\nonumber \\
& = & - \frac{1}{2} \frac{\partial}{\partial \mu} \int \! \frac{d\mathbf{k}}{(2 \pi)^3} \, T \, \sum_{n} G_{0}(k) G_{0}(-k) = - \frac{1}{2} \frac{\partial}{\partial \mu} R_{pp}(Q=0)
\nonumber \\
& \simeq & - \frac{1}{2} \frac{(2 m)^{3/2} }{8 \pi^{2} \sqrt{\mu}} \left\{ \left[ \ln \left( \frac{8 \mu e^{\gamma}}{\pi T} \right) - 2 \right] + 2 \right\} \simeq \frac{m^{2}}{8 \pi a_{F} k_{\mu}^{2}} \, .
\label{Popov-second-approximation} 
\end{eqnarray}
\end{small}

\noindent
To obtain this result, we have made use of the definition (\ref{bubble-pp-exact}) for the regularized particle-particle bubble (with $Q=0$) and of its approximate expression (\ref{approximate-BCS-integral}) valid in the BCS limit, as well as of the result $\ln(\mu/T) \simeq - \pi/(2 k_{\mu} a_{F})$ obtained by the Thouless criterion (\ref{Thouless-criterion}) in the absence of the Popov correction by assuming that $T$ is of the order of the critical temperature $T_{c}$.

Entering the results (\ref{Popov-first-approximation}) and (\ref{Popov-second-approximation}) in Eq.~(\ref{Popov-self-energy-definition}) with $Q=0$, we obtain eventually in the BCS limit
\begin{equation}
\Sigma_{\mathrm{Popov}}^{\mathrm{B}} \simeq \frac{m k_{\mu}}{6 \pi^{2}} = \frac{1}{3} \frac{(2m)^{3/2} \sqrt{\mu}}{4 \pi^{2}} \, .
\label{Popov-self-energy-approximate-BCS}
\end{equation}
\noindent
Here, the expression on the right-hand side makes it evident the presence of a factor $1/3$, which is required to eliminate the ``spurious'' factor $e^{-1/3}$ in the expression of the critical temperature as noted after Eq.~(\ref{mu-BCS-limit}).

To this end, we modify the original Thouless criterion (\ref{Thouless-criterion}) by including the Popov contribution, in the form:
\begin{equation}
\Gamma_{0}(Q=0;T_{c},\mu_{c})^{-1} - \Sigma_{\mathrm{Popov}}^{\mathrm{B}}(Q=0) = 0 \, .
\label{Thouless-criterion-Popov}
\end{equation}
With the help of the expressions (\ref{Thouless-criterion}) and (\ref{Popov-self-energy-approximate-BCS}), this modified Thouless criterion then yields the result
\begin{equation}
\frac{m}{4 \pi a_{F}} + \frac{(2m)^{3/2} \sqrt{\mu}}{4 \pi^{2}} \, \ln \left( \frac{8 \mu e^{\gamma} e^{1/3}}{\pi T_{c} e^{2}} \right) = 0
\label{Thouless-criterion-Popov-implemented}
\end{equation}
\noindent
as we had anticipated.

In the BEC (strong-coupling) limit, on the other hand, the Popov bosonic-like self-energy acquires the form \cite{Pieri-2005}:
\begin{equation}
\Sigma_{\mathrm{Popov}}^{\mathrm{B}} \simeq - \frac{m k_{F}}{6 \pi^{2}} \, \left( k_{F} a_{F} \right)^{2} \, .
\label{Popov-self-energy-approximate-BEC}
\end{equation}

In the following, the Popov correction (\ref{Popov-self-energy-definition}) (with $Q=0$) will be included numerically throughout the whole BCS-BEC crossover, 
where it will turn out to give an important contribution to the calculation of the critical temperature.
In this context, recovering numerically the analytic BCS [Eq.~(\ref{Popov-self-energy-approximate-BCS})] and BEC [Eq.~(\ref{Popov-self-energy-approximate-BEC})] limiting values 
will serve as an important test for the accuracy of the calculations. 
The procedure to set up the numerical calculation of $\Sigma_{\mathrm{Popov}}^{\mathrm{B}}$ for a generic value of the coupling $(k_{F}a_{F})^{-1}$ is described in  Appendix~\ref{sec:appendix-A}.

\vspace{0.05cm}
\begin{center}
{\bf D. GMB contribution}
\end{center}
\vspace{-0.2cm}

Akin to the Popov correction discussed above, also the GMB correction can be interpreted in terms of a bosonic-like self-energy $\Sigma_{\mathrm{GMB}}^{\mathrm{B}}$ shown in Fig.~\ref{Figure-2}(b), 
which dresses the bare pair propagator $\Gamma_{0}$.
In this respect, our approach is similar to the original GMB treatment \cite{GMB-1961}, where the singularities of the ``vertex part'' of the two-particle Green's function \cite{AGD-1963} 
were identified by approaching $T_{c}$ from the normal phase.
This contrasts with the treatment of Ref.~\cite{Heiselberg-2000} (see also Ref.~\cite{PS-2008}), where an induced interaction was added to the bare interaction directly in the linearised gap equation to approach $T_{c}$ from the superfluid phase.
In these references, however, only the BCS (weak-coupling) limit was considered.

The present treatment of the GMB correction differs from all previous treatments on the same subject.
This is because we shall consider the GMB correction not only in the BCS limit but also throughout the BCS-BEC crossover.
And, contrary to previous treatments of the GMB correction where the BCS-BEC crossover was considered \cite{Yu-2009,Ruan-2013}, \emph{we shall take into account the full dependence on wave vector and frequency of the pair propagators} $\Gamma_{0}$ that appear in Fig.~\ref{Figure-2}(b).
This dependence, in fact, will prove to be an essential ingredient for an appropriate description of the physics at the basis of the GMB correction.

The analytic expression of the GMB bosonic-like self-energy $\Sigma_{\mathrm{GMB}}^{\mathrm{B}}$ depicted in Fig.~\ref{Figure-2}(b) reads:
\begin{eqnarray}
& & \hspace{-0.5cm} \Sigma_{\mathrm{GMB}}^{\mathrm{B}}(Q) = \int \! \frac{d\mathbf{k}}{(2 \pi)^3} \, T \, \sum_{n} \int \! \frac{d\mathbf{p}}{(2 \pi)^3} \, T \sum_{\nu_{p}} \,
                                                                                                                                                                 \int \! \frac{d\mathbf{q}}{(2 \pi)^3} \, T \sum_{\nu_{q}} 
\nonumber \\
& \times & G_{0}(p-k+Q) \, G_{0}(k-p) \, G_{0}(p+q-k+Q) \, G_{0}(k) 
\nonumber \\
& \times & G_{0}(q-k+Q) \, G_{0}(k-q) \, \Gamma_{0}(p+Q) \, \Gamma_{0}(q+Q) \, .
\label{GMB-self-energy-definition}
\end{eqnarray}
\noindent
In the following, we shall again limit ourselves to consider the case $Q=0$, whereby we set $\Sigma_{\mathrm{GMB}}^{\mathrm{B}}=\Sigma_{\mathrm{GMB}}^{\mathrm{B}}(Q=0)$.

The expression (\ref{GMB-self-energy-definition}), which contains six single-particle propagators $G_{0}$  and two pair propagators $\Gamma_{0}$ with one fermionic and two bosonic four-vector integrations, is considerably
more involved than the Popov counterpart (\ref{Popov-self-energy-definition}), so that its numerical calculation for $Q=0$ will also prove quite more challenging. 
The numerical strategies to deal with this complicated calculation will be outlined in subsection \ref{sec:numerical-results}-A and further discussed in more detail in
Appendix~\ref{sec:appendix-A}.
In this context, we note that, for the needs of the numerical calculations, the expression (\ref{GMB-self-energy-definition}) is fully symmetric under the interchange $p \leftrightarrow q$.
Different sets of four-wave-vector integration/summation variables $(k,p,q)$, however, will be useful to obtain the BCS and BEC limits of the expression (\ref{GMB-self-energy-definition}), to be considered next.

In addition, we anticipate that in the numerical calculations reported in Section \ref{sec:numerical-results} the pair propagator $\Gamma_{0}$, which enters the Popov [Eq.~(\ref{Popov-self-energy-definition})] and GMB [Eq.~(\ref{GMB-self-energy-definition})] bosonic-like self-energies as well as the fermionic self-energy (\ref{Sigma-NSR}) used in the density equation (\ref{density}), will everywhere be replaced by the pair propagator $\Gamma$ dressed by a constant shift according to Eq.~(\ref{Gamma-1}) below.
The presence of this shift (either $\Sigma_{\mathrm{Popov}}^{\mathrm{B}}$ or $\Sigma_{\mathrm{GMB}}^{\mathrm{B}}$, or both) is required to avoid unwanted divergences, that would otherwise occur when $T \rightarrow T_{c}$.

\vspace{0.05cm}
\begin{center}
{\bf E. Disentanglement of pairing and screening \\ in the BCS limit}
\end{center}
\vspace{-0.2cm}

To obtain the BCS limiting value of the GMB bosonic-like self-energy (\ref{GMB-self-energy-definition}), it would appear natural to approximate both pair propagators $\Gamma_{0}$ 
therein by their asymptotic expression $-4 \pi a_{F}/m$ valid in the BCS limit, in a similar way to what we did in Eq.~(\ref{Popov-first-approximation}) for the Popov bosonic-like self-energy.
However, for the GMB case adopting \emph{tout court} this approximation for $\Gamma_{0}$ would make the GMB self-energy (\ref{GMB-self-energy-definition}) diverge, and 
some care should be accordingly exerted in this context.
\begin{figure}[t]
\begin{center}
\includegraphics[width=7.5cm,angle=0]{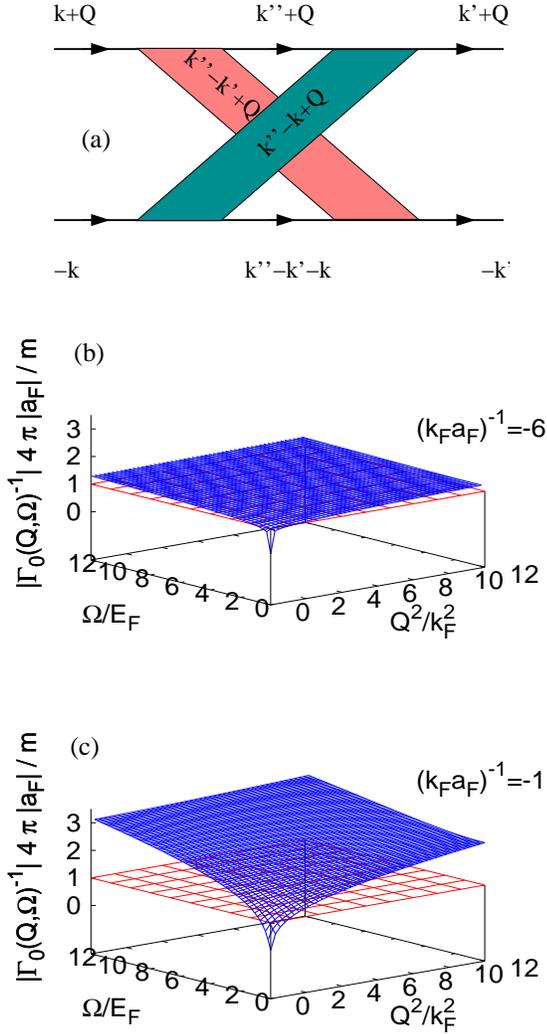}
\caption{(Color online) (a) The GMB bosonic-like self-energy $\Sigma_{\mathrm{GMB}}^{\mathrm{B}}$ of Fig.~\ref{Figure-2}(b) is cast in a form suitable to obtain its BCS limit
                                          analytically.
                                          The magnitude of the inverse $\Gamma_{0}(\mathbf{Q},\Omega_{\nu})^{-1}$ of the pair propagator (in units of $m/(4 \pi |a_{F}|)$) is shown vs 
                                          $(|\mathbf{Q}|/k_{F})^{2}$ and $\Omega_{\nu}/E_{F}$ for the couplings (b) $(k_{F}a_{F})^{-1}=-6$ in the extreme BCS limit and 
                                          (c) $(k_{F}a_{F})^{-1}=-1$ at the boundary between the BCS and the crossover regions.}
\label{Figure-3}
\end{center} 
\end{figure}
For these reasons, it is relevant to spell out in detail the sequence of (approximate) steps to obtain the GMB result for $T_{c}$ \cite{GMB-1961}.
In this way, it will also be clear that these approximations cannot simply be extended to the whole BCS-BEC crossover, for which a more complete approach is instead required.

To deal with the BCS limit of the GMB self-energy (\ref{GMB-self-energy-definition}), it is convenient to rename the four-wave-vector variables like in Fig.~\ref{Figure-3}(a) (where, again, the case $Q=0$ will 
only be considered). 
If one would retain the full four-vector dependence of the two pair propagators $\Gamma_{0}$ appearing in this diagram, there would be no problem in the convergence of the 
corresponding sums and integrals over the fermionic four-vectors $(k,k',k'')$, but no analytic result could be obtained in this way.

In the weak-coupling limit, the GMB result can be derived from the general expression (\ref{GMB-self-energy-definition}) through the following steps:

\noindent
(i) Begin by taking each $\Gamma_{0}$ of the form $-4 \pi a_{F}/m$, independent of wave vector and frequency.
By doing this, a particle-hole bubble of the form
\begin{eqnarray}
& & \chi_{\mathrm{ph}}(k+k') = \int \! \frac{d\mathbf{k''}}{(2\pi)^3} \, T \, \sum_{n''} \, G_{0}(\mathbf{k''},\omega_{n''})
\nonumber \\
& \times & G_{0}(\mathbf{k''-k-k'},\omega_{n''}-\omega_{n}-\omega_{n'}) 
\nonumber \\
& = & \int \! \frac{d\mathbf{k''}}{(2 \pi)^3} \frac{f(\xi_{\mathbf{k}+\mathbf{k}'+\mathbf{k''}}) - f(\xi_{\mathbf{k''}})} 
{\xi_{\mathbf{k}+\mathbf{k}'+\mathbf{k''}}-\xi_{\mathbf{k''}}-i(\omega_{n}+\omega_{n'})}
\label{particle-hole-bubble}
\end{eqnarray}
\noindent
appears in the central part of the diagram.
This bubble identifies the simplest process associated with screening in a Fermi gas \cite{Mahan-2000}.

\noindent
(ii) In the particle-hole bubble (\ref{particle-hole-bubble}), set $\omega_{n}+\omega_{n'}=0$, take $\mathbf{k}$ and $\mathbf{k}'$ on a Fermi sphere with radius $k_{\mu}$ and average over their relative angle.
The result is
\begin{equation}
\bar{\chi}_{\mathrm{ph}} = -N(\mu) \ln (4e)^{1/3} \, 
\label{average-particle-hole-bubble}
\end{equation}
\noindent
where $N(\mu)=(2m)^{3/2}\sqrt{\mu}/(4 \pi^{2})$ is the single-particle density of states (per spin component) taken at the chemical potential.
This step is justified by the presence of the sums over the four-vectors $k$ and $k'$ in Fig.~\ref{Figure-3}(a), whereby the factors $G_{0}(k) G_{0}(-k)$ and $G_{0}(k') G_{0}(-k')$ in the particle-particle bubbles on the left and right sides of the diagram are strongly peaked at ($|\mathbf{k}|=k_{\mu}, \omega_{n}=0)$ and ($|\mathbf{k'}|=k_{\mu}, \omega_{n'}=0)$.

\noindent
(iii) In this way, the diagram of Fig.~\ref{Figure-3}(a) is effectively \emph{disentangled} into the product of three terms, namely, two particle-particle bubbles of the form 
(\ref{particle-particle-bubble}) that appear on the left and right sides of the diagram and the (averaged) particle-hole bubble that appears in the central part of the diagram.
Taken as they stand, the two particle-particle bubbles would diverge in the ultraviolet.
However, owing to the convergence of the original diagram where the two pair propagators $\Gamma_{0}$ retain their full wave-vector and frequency dependence,
one can safely make the integrals over $\mathbf{k}$ and $\mathbf{k'}$ convergent again by replacing each particle-particle bubble (\ref{particle-particle-bubble}) with its regularized version (\ref{bubble-pp-exact}),
whose integrand has the same behavior in the dominant region where ($|\mathbf{k}| \simeq k_{\mu}, \omega_{n} \simeq 0)$ and ($|\mathbf{k'}| \simeq k_{\mu}, \omega_{n'} \simeq 0)$.
Accordingly, the GMB self-energy becomes:
\begin{equation}
\Sigma_{\mathrm{GMB}}^{\mathrm{B}} = \Sigma_{\mathrm{GMB}}^{\mathrm{B}}(Q=0) \simeq \! \left(\! - \frac{4 \pi a_{F}}{m} \!\right)^{2} \!R_{pp}(Q=0)^{2} \bar{\chi}_{\mathrm{ph}} \, .
\label{GMB-first-approximation-BCS}
\end{equation}
\noindent
(Note in this context that the use here of the word ``disentanglement'' follows the original Feynman's work \cite{Feynman-1951}.)

\noindent
(iv) The approximate result (\ref{GMB-first-approximation-BCS}) affects the critical temperature by modifying the Thouless criterion, in the form
\begin{equation}
\Gamma_{0}(Q=0;T_{c},\mu_{c})^{-1} - \Sigma_{\mathrm{GMB}}^{\mathrm{B}}(Q=0) = 0 
\label{Thouless-criterion-GMB}
\end{equation}
\noindent
in analogy with Eq.~(\ref{Thouless-criterion-Popov}) for the Popov case.
Making use of the expression (\ref{Gamma-0}) for $\Gamma_{0}$ and of the result (\ref{GMB-first-approximation-BCS}) for $\Sigma_{\mathrm{GMB}}^{\mathrm{B}}$, the generalised Thouless
criterion (\ref{Thouless-criterion-GMB}) becomes:
\begin{equation}
\frac{m}{4 \pi a_{F}} + R_{pp}(Q=0) + \left( \frac{4 \pi a_{F}}{m} R_{pp}(Q=0) \right)^{2} \bar{\chi}_{\mathrm{ph}} = 0 \, .
\label{Thouless-criterion-GMB-approximate}
\end{equation}

\noindent
(v) The expression (\ref{Thouless-criterion-GMB-approximate}) can be further simplified by noting that, for $T$ of the order of the BCS critical temperature, the Thouless criterion (\ref{Thouless-criterion}) 
is equivalent to writing $\frac{4 \pi a_{F}}{m} R_{pp}(Q=0) = -1$.
Using this result, Eq.~(\ref{Thouless-criterion-GMB-approximate}) reduces to its final form:
\begin{equation}
\frac{m}{4 \pi a_{F}} + R_{pp}(Q=0) + \bar{\chi}_{\mathrm{ph}} = 0 \, .
\label{Thouless-criterion-GMB-approximate-final}
\end{equation}
\noindent
With the result (\ref{average-particle-hole-bubble}), one then obtains in analogy with Eq.~(\ref{Thouless-criterion-Popov-implemented}):
\begin{equation}
\frac{m}{4 \pi a_{F}} + \frac{(2m)^{3/2} \sqrt{\mu}}{4 \pi^{2}} \, \ln \left( \frac{8 \mu e^{\gamma}}{\pi T_{c} e^{2} (4 e)^{1/3}} \right) = 0 \, .
\label{Thouless-criterion-GMB-implemented}
\end{equation}
\noindent
Upon setting $\mu=E_{F}$ in this expression, one gets that the value of the BCS critical temperature (\ref{Tc-BCS-no_cutoff}) is reduced by the factor $(4 e)^{1/3} \simeq 2.2$, as it was obtained in Ref.\cite{GMB-1961}.
Otherwise, if the value (\ref{mu-BCS-limit}) of $\mu$ within the non-self-consistent t-matrix approximation is adopted, the Popov contribution is also needed to get rid of the additional ``spurious'' factor $e^{-1/3}$ as discussed in subsection \ref{sec:G-MB-BCS-BEC}-C.

From the way it has been derived, it is clear that Eq.~(\ref{Thouless-criterion-GMB-approximate-final}) holds only in the \emph{extreme} BCS (weak-coupling) limit $(k_{F}a_{F})^{-1} \ll -1$, whereby
$\Gamma_{0}(Q)$ can be approximated by the constant term $- 4 \pi a_{F}/m$.
We have explicitly verified the validity of this approximation numerically, by plotting the magnitude of the inverse $\Gamma_{0}(\mathbf{Q},\Omega_{\nu})^{-1}$ of the pair propagator 
in units of $m/(4 \pi |a_{F}|)$, in Fig.~\ref{Figure-3}(b) for the coupling values $(k_{F}a_{F})^{-1}=-6$ in the extreme BCS (weak-coupling) limit and in Fig.~\ref{Figure-3}(c) for the coupling 
value $(k_{F}a_{F})^{-1}=-1$ at the boundary between the BCS and the crossover regions.
From these plots, one indeed verifies that in the extreme BCS (weak-coupling) limit to a good approximation $\Gamma_{0}(\mathbf{Q},\Omega_{\nu})$ can be considered constant over a large plateau in the $(|\mathbf{Q}|,\Omega_{\nu})$ plane.
This property, however, no longer holds already at the boundary between the BCS and the crossover regions, where strong deviations of $\Gamma_{0}$ from constancy appear evident.
In spite of this property, in Refs.~\cite{Yu-2009,Ruan-2013} the critical temperature was calculated by relying on the result (\ref{Thouless-criterion-GMB-approximate-final}) not only in the BCS limit but also across the whole BCS-BEC crossover.

In the following, we shall remedy this shortcoming by maintaining the full $\mathbf{Q}$ and $\Omega_{\nu}$ dependence of $\Gamma_{0}$ in the expression (\ref{GMB-self-energy-definition})
of the GMB self energy $\Sigma_{\mathrm{GMB}}^{\mathrm{B}}$.
The corresponding numerical calculation will be reported in Section~\ref{sec:numerical-results} throughout the BCS-BEC crossover.
In particular, we anticipate that our calculation will be able to reproduce in a totally numerical fashion the GMB result for the reduction of the value of the critical temperature (\ref{Tc-BCS-no_cutoff}) 
by the factor $(4 e)^{1/3} \simeq 2.2$ in the extreme weak-coupling limit, thus confirming the validity of the non-trivial weak-coupling approximations leading to the GMB result.

\vspace{0.05cm}
\begin{center}
{\bf F. Transmuting of screening into pairing \\ in the BEC limit}
\end{center}
\vspace{-0.2cm}

\begin{figure}[t]
\begin{center}
\includegraphics[width=7.5cm,angle=0]{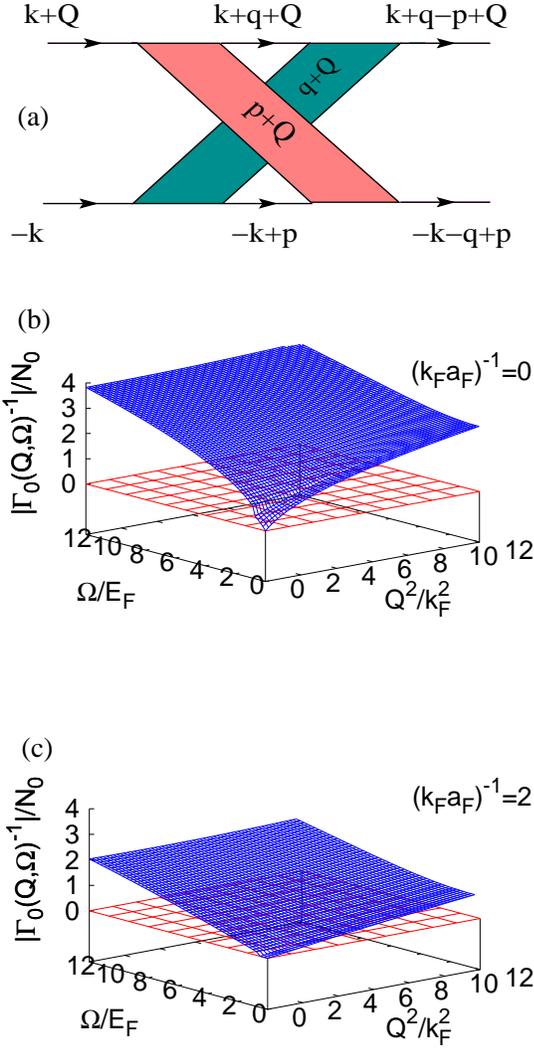}
\caption{(Color online) (a) The GMB bosonic-like self-energy $\Sigma_{\mathrm{GMB}}^{\mathrm{B}}$ of Fig.~\ref{Figure-2}(b) is cast in a form suitable to obtain its BEC limit
                                          analytically.
                                          The magnitude of the inverse $\Gamma_{0}(\mathbf{Q},\Omega_{\nu})^{-1}$ of the pair propagator (in units of $N_{0}=N(\mu=E_{F})$) is shown vs 
                                          $(|\mathbf{Q}|/k_{F})^{2}$ and $\Omega_{\nu}/E_{F}$ for the couplings (b) $(k_{F}a_{F})^{-1}=0$ at unitarity and 
                                          (c) $(k_{F}a_{F})^{-1}=2$ in the BEC regime.}
\label{Figure-4}
\end{center} 
\end{figure}

To deal with the BEC (strong-coupling) limit of the GMB bosonic-like self-energy (\ref{GMB-self-energy-definition}), it is convenient to rename the four-wave-vector variables like in Fig.~\ref{Figure-4}(a) (where the case $Q=0$ is again considered).
In this limit, we are going to show analytically that keeping the full $\mathbf{Q}$ and $\Omega_{\nu}$ dependence in the pair propagators $\Gamma_{0}$ of the GMB contribution is \emph{essential} for a correct evaluation of this quantity.
This need can be anticipated by looking at the much stronger $\mathbf{Q}$ and $\Omega_{\nu}$ dependence of $|\Gamma_{0}(\mathbf{Q},\Omega_{\nu})^{-1}|$ that occurs in the BEC with respect to the BCS limit, as shown in Fig.~\ref{Figure-4}(c).
For comparison, in Fig.~\ref{Figure-4}(b) the shape of $|\Gamma_{0}(\mathbf{Q},\Omega_{\nu})^{-1}|$ is also shown at unitarity.
[Note the change of normalization for $|\Gamma_{0}(\mathbf{Q},\Omega_{\nu})^{-1}|$ in Figs.~\ref{Figure-4}(b) and \ref{Figure-4}(c) with respect to Figs.~\ref{Figure-3}(b) and \ref{Figure-3}(c).]

In the BEC limit $\mu/T \rightarrow - \infty$ we are interested in, the pair propagators entering the diagram of Fig.~\ref{Figure-4}(a) have the approximate form \cite{PS-2000}:
\begin{equation}
\Gamma_{0}(\mathbf{q},\Omega_{q}) = \frac{D(\mathbf{q},\Omega_{q})}{i \Omega_{q} - \xi^{B}_{\mathbf{q}}}
\label{pair-propagator-BEC-limit}
\end{equation}
\noindent
where
\begin{equation}
D(\mathbf{q},\Omega_{q}) = - \frac{4 \pi}{m^{2} a_{F}} \, \left( 1 \, + \, \sqrt{1 \, + \, \frac{\xi^{B}_{\mathbf{q}} - i \Omega_{q}}{\epsilon_{0}} } \right) \, .
\label{definition-D}
\end{equation}
\noindent
In these expressions, $\epsilon_{0} = (m a_{F}^{2})^{-1}$ is the binding energy of the two-fermion problem in vacuum and 
$\xi^{B}_{\mathbf{q}} = \mathbf{q}^{2}/(4m) - \mu_{B}$ where $\mu_{B} = 2\mu + \epsilon_{0}$ is the chemical potential for the composite bosons that form in this limit.
When extended to the complex frequency plane by letting $ i \Omega_{q} \rightarrow z$ in Eqs.~(\ref{pair-propagator-BEC-limit}) and (\ref{definition-D}), the function $\Gamma_{0}(\mathbf{q},z)$ has a pole at $z = \xi^{B}_{\mathbf{q}}$ and a cut along the real frequency axis for $z \ge \xi^{B}_{\mathbf{q}} + \epsilon_{0}$.

The integrations occurring in the expression of the GMB bosonic-like self-energy $\Sigma_{\mathrm{GMB}}^{B}$ (cf. the diagram of Fig.~\ref{Figure-4}(a)) can be done by first considering
the sums over the bosonic Matsubara frequencies $\Omega_{p}$ and $\Omega_{q}$. 
These sums can both be transformed into contour integrals in the complex $z$-plane by using the bosonic distribution $b(z) = (\exp\{z/(k_{B}T)\} -1)^{-1}$ \cite{FW}.
Let $C$ be a contour in this plane which encircles poles and branch cut clockwise.
Owing to the presence of the function $b(z)$, the contribution from the cuts of $\Gamma_{0}(p)$ and $\Gamma_{0}(q)$ are exponentially suppressed in the BEC limit (when $\epsilon_{0}$ is the largest energy scale) and need not be considered.
Similarly, the contribution from the poles of $G_{0}(k+q)$ and $G_{0}(-k+p)$ where $p$ and $q$ appear with a positive sign are also exponentially suppressed.
Therefore, only the poles of $\Gamma_{0}(p)$ and $\Gamma_{0}(q)$ and of $G_{0}(k+q-p)$ and $G_{0}(-k-q+p)$ (where $p$ or $q$ appear with a negative sign) give a finite contribution.

Specifically, three contributions result by doing the sums over $\Omega_{q}$  and $\Omega_{p}$ in the sequence, which arise respectively from the poles of:
(i) $\Gamma_{0}(q)$ and $\Gamma_{0}(p)$;
(ii) $\Gamma_{0}(q)$ and $G_{0}(k+q-p)$;
(iii) $\Gamma_{0}(p)$ and $G_{0}(-k-q+p)$.

\vspace{0.1cm}
The contribution (i) yields:
\begin{eqnarray}
& & \Sigma_{\mathrm{GMB}}^{B \mathrm{(i)}} \cong \frac{8 \pi}{m^{2} a_{F}} \int \! \frac{d\mathbf{q}}{(2 \pi)^{3}} \, b(\xi^{B}_{\mathbf{q}})
\nonumber \\
& \times & \frac{8 \pi}{m^{2} a_{F}} \int \! \frac{d\mathbf{p}}{(2 \pi)^{3}} \, b(\xi^{B}_{\mathbf{p}}) \, \int \! dk \, G_{0}(k)^{3} G_{0}(-k)^{3} 
\label{contribution-i}
\end{eqnarray}
\noindent
where
\begin{equation}
\int \! dk \, G_{0}(k)^{3} G_{0}(-k)^{3} \cong \frac{15 m^{5} a_{F}^{7}}{256 \pi} 
\label{G-0-3}
\end{equation}
\noindent
with the short-hand notation
\begin{equation}
\int \! dk = \int \! \frac{d\mathbf{k}}{(2 \pi)^{3}} \, T \, \sum_{\omega_{k}} 
\label{convention-four-integrals}
\end{equation}
\noindent
and similarly for the other integrals.
Note that this contribution (which is of second order in the bosonic density $n_{B} = n/2$) could have been obtained by neglecting the $p$ and $q$ dependence everywhere in the $G_{0}$ of the diagram in Fig.~\ref{Figure-4}(a) and by approximating further:
\begin{eqnarray}
\int \! dq \, \Gamma_{0}(q) & \cong & - \frac{8 \pi}{m^{2} a_{F}} \, \int \! \frac{d\mathbf{q}}{(2 \pi)^{3}} \, T \, \sum_{\Omega_{q}} \,
\frac{e^{i \Omega_{q} \eta}}{i \Omega_{q} - \xi^{B}_{\mathbf{q}}}
\nonumber \\
& = & \frac{8 \pi}{m^{2} a_{F}} \int \! \frac{d\mathbf{q}}{(2 \pi)^{3}} \, b(\xi^{B}_{\mathbf{q}}) = \frac{8 \pi}{m^{2} a_{F}} \, n_{B} 
\label{approximate-bosonic-density}
\end{eqnarray}
\noindent
where $\eta$ is a positive infinitesimal.

The contribution (ii) yields instead:
\begin{equation}
\Sigma_{\mathrm{GMB}}^{B \mathrm{(ii)}} \simeq \frac{8 \pi}{m^{2} a_{F}} \! \int \! \frac{d\mathbf{q}}{(2 \pi)^{3}} \, b(\xi^{B}_{\mathbf{q}})
\int \! \frac{d\mathbf{p}}{(2 \pi)^{3}} \, \phi(\mathbf{p}) 
\label{contribution-ii}
\end{equation}
\noindent
where the sum over the fermionic frequency $\omega_{k}$ has been performed analytically and with the definition 
\begin{equation}
\phi(\mathbf{p}) = \int \! \frac{d\mathbf{k}}{(2 \pi)^{3}} \, 
\frac{\Gamma_{0}(\mathbf{p},- \xi_{\mathbf{k}} - \xi_{\mathbf{k}-\mathbf{p}})}{(2 \, \xi_{\mathbf{k}})^{2} (2 \, \xi_{\mathbf{k}-\mathbf{p}})^{2}} \, .
\label{function-phi}
\end{equation}
\noindent
Here, $\Gamma_{0}(\mathbf{p},- \xi_{\mathbf{k}} - \xi_{\mathbf{k}-\mathbf{p}})$ is obtained from $\Gamma_{0}(\mathbf{p},\Omega_{p})$ given by 
Eqs.~(\ref{pair-propagator-BEC-limit}) and (\ref{definition-D}) with the replacement $i \Omega_{p} \rightarrow - \xi_{\mathbf{k}} - \xi_{\mathbf{k}-\mathbf{p}}$.
Note that, the numerator (\ref{definition-D}) of the expression (\ref{pair-propagator-BEC-limit}) contributes to the integration in Eq.~(\ref{function-phi}).
It is for this reason that, contrary to the expression (\ref{contribution-i}), only one power of the bosonic density $n_{B}$ appears eventually in the expression (\ref{contribution-ii}).
An expression identical to (\ref{contribution-ii}) results also from the contribution (iii) (apart from the interchange of the wave vectors $\mathbf{q}$ and $\mathbf{p}$).

There remains to calculate the function $\phi(\mathbf{p})$ given by Eq.~(\ref{function-phi}).
To this end, it is convenient to change the integration variable into $\mathbf{k}' = \mathbf{k} - \mathbf{p}/2$, rescale the magnitude of the wave vectors by
$\tilde{k}' = |\mathbf{k'}| a_{F}$ and $\tilde{p} = |\mathbf{p}| a_{F}$, and introduce the notation
\begin{equation}
A(\tilde{k}',\tilde{p}) = \frac{1 + \tilde{k}'^{2} + \tilde{p}^{2}/4 }{\tilde{k}' \, \tilde{p}} \, .
\label{A-definition}
\end{equation}
\noindent
In this way, the integral over the angle between $\mathbf{k}'$ and $\mathbf{p}$ can be done analytically, yielding:
\begin{equation}
\phi(\mathbf{p}) = \frac{2 \, m^{3} \, a_{F}^{4}}{\pi} \, \frac{1}{\mathbf{p}^{2}} \, F(\tilde{p})
\label{function-phi-final}
\end{equation}
\noindent
where
\begin{eqnarray}
F(\tilde{p}) & = & \int_{0}^{\infty} \! d \tilde{k} \, \frac{\left( 1 \, + \, \sqrt{2 + \tilde{k}^{2} + \tilde{p}^{2}/2 }\right)}
{\left(  2 + 2 \tilde{k}^{2} + \tilde{p}^{2} \right) \left(  1 + \tilde{k}^{2} + \tilde{p}^{2}/4 \right)^{2}}
\label{F-definition} \\
& \times & \left[  \frac{1}{A(\tilde{k},\tilde{p})^{2} - 1} \, + \, \frac{1}{A(\tilde{k},\tilde{p})} \, \mathrm{arctanh}\left(\frac{1}{A(\tilde{k},\tilde{p})}\right) \right] \, .
\nonumber
\end{eqnarray}
\noindent
To obtain this expression, we have neglected the small energy scale $\mu_{B}$ with respect to $\epsilon_{0}$.
We can thus rewrite in Eq.~(\ref{contribution-ii}):
\begin{equation}
\int \! \frac{d\mathbf{p}}{(2 \pi)^{3}} \, \phi(\mathbf{p}) = \frac{m^{3} \, a_{F}^{3}}{\pi^{3}} \,  \int_{0}^{\infty} \! d \tilde{p} \, F(\tilde{p}) 
\label{function-phi-integral}
\end{equation}
\noindent
where a numerical calculation gives the value $\tilde{I} = 0.25974$ for the integral of $F(\tilde{p})$ in Eq.~(\ref{function-phi-integral}).

In conclusion, the sum of the above three contributions (i)-(iii) can be written in the compact form:
\begin{eqnarray}
\Sigma_{\mathrm{GMB}}^{B} & \simeq &
\int \! \frac{d\mathbf{p}}{(2 \pi)^{3}} \int \! \frac{d\mathbf{q}}{(2 \pi)^{3}} 
\nonumber \\
& \times & \left\{ \frac{8 \pi}{m^{2} a_{F}} \left[ b(\xi^{B}_{\mathbf{q}}) \, \phi(\mathbf{p}) \, + \, b(\xi^{B}_{\mathbf{p}}) \, \phi(\mathbf{q}) \right] \right.
\nonumber \\
& + & \left. \left( \frac{8 \pi}{m^{2} a_{F}} \right)^{2} \, \frac{15 m^{5} a_{F}^{7}}{256 \pi} \, b(\xi^{B}_{\mathbf{p}}) \, b(\xi^{B}_{\mathbf{q}}) \right\}
\nonumber \\
& \simeq & \frac{16 \, \tilde{I}}{\pi^{2}} \, \frac{m k_{F}}{6 \pi^{2}} \left( k_{F} a_{F} \right)^{2} 
\label{GMB-approximation-BEC}
\end{eqnarray}
\noindent
to the leading order in the small parameter $k_{F} a_{F}$.
Apart from a sign, this result differs from the corresponding Popov result (\ref{Popov-self-energy-approximate-BEC}) by the constant factor $16 \, \tilde{I}/\pi^{2} \simeq 0.421$.

Note that the power-law dependence of the expression (\ref{GMB-approximation-BEC}) on the small parameter $(k_{F} a_{F}) \ll1$ contrasts with the exponential dependence
of the particle-hole bubble (\ref{particle-hole-bubble}) on the (square of the) coupling parameter $(k_{F} a_{F})^{-1}$, which would be obtained by extending the expression (\ref{particle-hole-bubble}) to the BEC limit $(k_{F} a_{F})^{-1} \rightarrow + \infty$.
In fact, no trace of the particle-hole bubble (\ref{particle-hole-bubble}) appears in the derivation of the result (\ref{GMB-approximation-BEC}), where only particle-particle processes, which are relevant to the scattering between composite bosons, correctly occur in the BEC limit.
This result confirms that screening processes show up in the GMB contribution \emph{only\/} in the opposite BCS limit $(k_{F} a_{F})^{-1} \rightarrow - \infty$ treated originally by Gor'kov and Melik-Barkhudarov \cite{GMB-1961} and discussed in subsection~\ref{sec:G-MB-BCS-BEC}-E.

The result (\ref{GMB-approximation-BEC}) will be further considered in Appendix~\ref{sec:appendix-B}, where the GMB self-energy $\Sigma_{\mathrm{GMB}}^{B}$ is shown to contribute to the scattering length $a_{B}$ of composite bosons that form in the BEC limit.
In the following Section, the result (\ref{GMB-approximation-BEC}) will instead represent a benchmark for the numerical calculation in the (extreme) BEC limit.

\section{Numerical results} 
\label{sec:numerical-results}

In this Section, we implement numerically the inclusion of the Popov and GMB corrections into the non-self-consistent t-matrix approximation.
We shall specifically be concerned with determining how these corrections affect the value of the critical temperature $T_{c}$ throughout the BCS-BEC crossover.
In the process, the numerical accuracy of the calculation of the Popov and GMB corrections will be tested against the analytic results obtained in Section~\ref{sec:G-MB-BCS-BEC} in the BCS and BEC limits.
Our numerical results for $T_{c}$ will also be compared with Quantum Monte Carlo calculations which are available in an extended region of coupling about unitarity and with the experimental data which are available at unitarity.

\vspace{0.05cm}
\begin{center}
{\bf A. Numerical strategies at $T_{c}$}
\end{center}
\vspace{-0.2cm}

In the Popov ($\Sigma_{\mathrm{Popov}}^{B}$ of Eq.~(\ref{Popov-self-energy-definition})) and GMB ($\Sigma_{\mathrm{GMB}}^{B}$ of Eq.~(\ref{GMB-self-energy-definition})) bosonic-like self-energies depicted in Fig.~\ref{Figure-2}, all pair propagators were taken to be bare ones.
This choice was sufficient for obtaining the analytic results in the BCS and BEC limit.
When implementing the numerical calculations for these bosonic-like self-energies, however, care must be exerted on the fact that the bare pair propagator $\Gamma_{0}(Q)$ becomes singular
at $Q=0$ for specific values of $T_{c}$ and $\mu_{c}$ according to the Thouless criterion (\ref{Thouless-criterion}).
This problem is definitely bound to show up numerically in the BCS limit, where one knows that the GMB correction decreases the value of the critical temperature by a factor $2.2$ with respect to the BCS result.

To overcome this problem, it is clear that \emph{some degree of self-consistency} has unavoidably to be included in the calculation.
However, to keep at the same time the calculation for determining $T_{c}$ as simple as possible (thus avoiding to increase its complexity beyond affordable limits), self-consistency will be implemented in practice according to the following scheme.

We begin by considering Eq.~(\ref{Gamma-0}), in the form
\begin{equation}
\Gamma_{0}(Q;T,\mu)^{-1} = - \frac{m}{4 \pi a_{F}} - R_{\mathrm{pp}}(Q;T,\mu)
\label{Gamma-0-1}
\end{equation}
\noindent
where $R_{\mathrm{pp}}$ is given by the expression (\ref{bubble-pp-exact}) and the dependence on the thermodynamic quantities $(T,\mu)$ has been explicitly indicated.
By normalising $\Gamma_{0}^{-1}$ in terms of the single-particle density of states (per spin component) at the Fermi level $N_{0}=N(\mu=E_{F})= m k_{F}/(2 \pi^{2})$, the coupling
$(k_{F} a_{F})^{-1}$ is seen to appear explicitly only in the first term on the right-hand side of Eq.~(\ref{Gamma-0-1}).
In particular, for given value of $(k_{F} a_{F})^{-1}$ the condition $\Gamma_{0}(Q=0;T_{c},\mu_{c})^{-1}=0$ determines a set of values $\{T_{c},\mu_{c}\}$ consistently with the equation
\begin{equation}
\frac{m}{4 \pi a_{F}} + R_{\mathrm{pp}}(Q=0;T_{c},\mu_{c}) = 0  \, .
\label{Gamma-0-1-Q=0}
\end{equation}
\noindent
For each of these pairs of values, we can then rewrite Eq.~(\ref{Gamma-0-1}) in the form
\begin{equation}
\Gamma_{0}(Q;T_{c},\mu_{c})^{-1} = - R_{\mathrm{pp}}(Q;T_{c},\mu_{c}) +  R_{\mathrm{pp}}(Q=0;T_{c},\mu_{c}) \, .
\label{Gamma-0-1-Tc}
\end{equation}
\noindent
Upon entering the expression (\ref{Gamma-0-1-Tc}) for $\Gamma_{0}(Q)$ into the fermionic self-energy (\ref{Sigma-NSR}) and then into the density equation (\ref{density}), in order to determine uniquely a pair of values $T_{c}$ and $\mu_{c}$ for given $(k_{F} a_{F})^{-1}$, it turns out that the density equation does not depend \emph{explicitly} on the coupling $(k_{F} a_{F})^{-1}$ but only on the pair 
$(T_{c},\mu_{c})$. 

Next, we consider the effect of the bosonic-like self-energy $\Sigma^{B}$ (either $\Sigma_{\mathrm{Popov}}^{B}$ or $\Sigma_{\mathrm{GMB}}^{B}$, or both) calculated at $Q=0$ only, by introducing
the \emph{dressed pair propagator}:
\begin{equation}
\Gamma(Q;T,\mu)^{-1} = \Gamma_{0}(Q;T,\mu)^{-1} - \Sigma^{B}(T,\mu) \, .
\label{Gamma-1}
\end{equation}
\noindent
In this case, a new set of values $\{\bar{T}_{c},\bar{\mu}_{c}\}$ is determined by the generalised Thouless criterion
\begin{equation}
\Gamma(Q=0;\bar{T}_{c},\bar{\mu}_{c})^{-1} = \Gamma_{0}(Q=0;\bar{T}_{c},\bar{\mu}_{c})^{-1} - \Sigma^{B}(\bar{T}_{c},\bar{\mu}_{c}) = 0 \, .
\label{generalized-Thouless-criterion}
\end{equation}
\noindent
In this way, Eq.~(\ref{Gamma-1}) becomes for given value of $(\bar{T}_{c},\bar{\mu}_{c})$ and a generic value of $Q$:
\begin{eqnarray}
& & \Gamma(Q;\bar{T}_{c},\bar{\mu}_{c})^{-1} = \Gamma_{0}(Q;\bar{T}_{c},\bar{\mu}_{c})^{-1} - \Sigma^{B}(\bar{T}_{c},\bar{\mu}_{c}) 
\nonumber \\
& = & \Gamma_{0}(Q;\bar{T}_{c},\bar{\mu}_{c})^{-1} - \Gamma_{0}(Q=0;\bar{T}_{c},\bar{\mu}_{c})^{-1} 
\nonumber \\
& = &  - R_{\mathrm{pp}}(Q;\bar{T}_{c},\bar{\mu}_{c}) +  R_{\mathrm{pp}}(Q=0;\bar{T}_{c},\bar{\mu}_{c})
\label{Gamma-1-revised-Tc} 
\end{eqnarray}
\noindent
where in the last line Eq.~(\ref{Gamma-0-1}) has been used.
Note that the right-hand sides of the two equations (\ref{Gamma-0-1-Tc}) and (\ref{Gamma-1-revised-Tc}) are formally identical to each other, apart from the different set of values 
$(T_{c},\mu_{c})$ and $(\bar{T}_{c},\bar{\mu}_{c})$ on which they depend.
This remark implies that the replacement $\Gamma_{0}(Q)^{-1} \rightarrow \Gamma(Q)^{-1}$ in the expressions of the Popov self-energy (\ref{Popov-self-energy-definition})
and the GMB self-energy (\ref{GMB-self-energy-definition}) amounts to considering a new set of values $(\bar{T}_{c},\bar{\mu}_{c})$ in the place of the old ones $(T_{c},\mu_{c})$.
This is also true when this replacement is made in the density equation (\ref{density}), which in this way depends \emph{only} on the pairs $(\bar{T}_{c},\bar{\mu}_{c})$ but not explicitly on the coupling $(k_{F} a_{F})^{-1}$.
The new values $(\bar{T}_{c},\bar{\mu}_{c})$ are then determined by the generalized Thouless criterion (\ref{generalized-Thouless-criterion}), in the form
\begin{eqnarray}
& - & \Gamma_{0}(Q=0;\bar{T}_{c},\bar{\mu}_{c})^{-1} + \Sigma^{B}(\bar{T}_{c},\bar{\mu}_{c})
\label{generalized-Thouless-criterion-final-form} \\
& = & \frac{m}{4 \pi a_{F}} + R_{\mathrm{pp}}(Q=0;\bar{T}_{c},\bar{\mu}_{c}) + \Sigma^{B}(\bar{T}_{c},\bar{\mu}_{c}) = 0 
\nonumber 
\end{eqnarray}
for a given coupling $(k_{F} a_{F})^{-1}$.
Note that Eq.~(\ref{generalized-Thouless-criterion-final-form}) has the same structure of Eq.~(\ref{Thouless-criterion-GMB}), although with a more general expression for the bosonic-like self energy $\Sigma^{B}$.

We again emphasize that the above simplified procedure, for including some degree of self-consistency in the pair propagator $\Gamma_{0}$, relies on the fact that we are considering only the $Q=0$ value of $\Sigma^{B}(Q)$ and limit ourselves to determine the critical temperature $T_{c}$.
In addition, we note that, owing to the formal analogy between Eqs.~(\ref{Gamma-0-1-Tc}) and (\ref{Gamma-1-revised-Tc}), the density equation (\ref{density}) (with the dressed $\Gamma$ in the place of the bare $\Gamma_{0}$) is formally identical to its counterpart within the (non-self-consistent) t-matrix approximation.
As a consequence, the solution of the generalised Thouless criterion (\ref{generalized-Thouless-criterion-final-form}) plus the associated density equation to get a new pair of values 
$(\bar{T}_{c},\bar{\mu}_{c})$, amounts \emph{in practice} to fixing an old pair of values $(T_{c},\mu_{c})$ that satisfy the original Thouless criterion (\ref{Gamma-0-1-Q=0}) plus the corresponding density equation
for given coupling $g=(k_{F} a_{F})^{-1}$, and then finding the value of the modified coupling $\bar{g}$ for which Eq.~(\ref{generalized-Thouless-criterion-final-form}) is satisfied.

Finally, we comment that more sophisticated degrees of self-consistency with respect to the one adopted here (like the replacement of the bare fermionic single-particle propagator $G_{0}$ by the dressed one $G$, everywhere $G_{0}$ appears in the relevant diagrams) are bound to result in an exceedingly difficult numerical calculation, especially as far as the GMB self-energy $\Sigma_{\mathrm{GMB}}^{B}$ is concerned.
This need to restrict to the $G_{0}$ will be apparent in the discussion of Appendix~\ref{sec:appendix-A}, where the calculation of the Popov and GMB diagrams will be implemented in detail.

\vspace{0.05cm}
\begin{center}
{\bf B. Bosonic-like self-energies and generalised Thouless criterion}
\end{center}
\vspace{-0.2cm}

To solve the generalized Thouless criterion (\ref{generalized-Thouless-criterion-final-form}), knowledge is required of the bosonic-like self-energy (either Popov or GMB, or both), calculated 
at the self-consistent values $(\bar{T}_{c},\bar{\mu}_{c})$ for a given value of the coupling $(k_{F}a_{F})^{-1}$.

\begin{figure}[t]
\begin{center}
\includegraphics[width=7.0cm,angle=0]{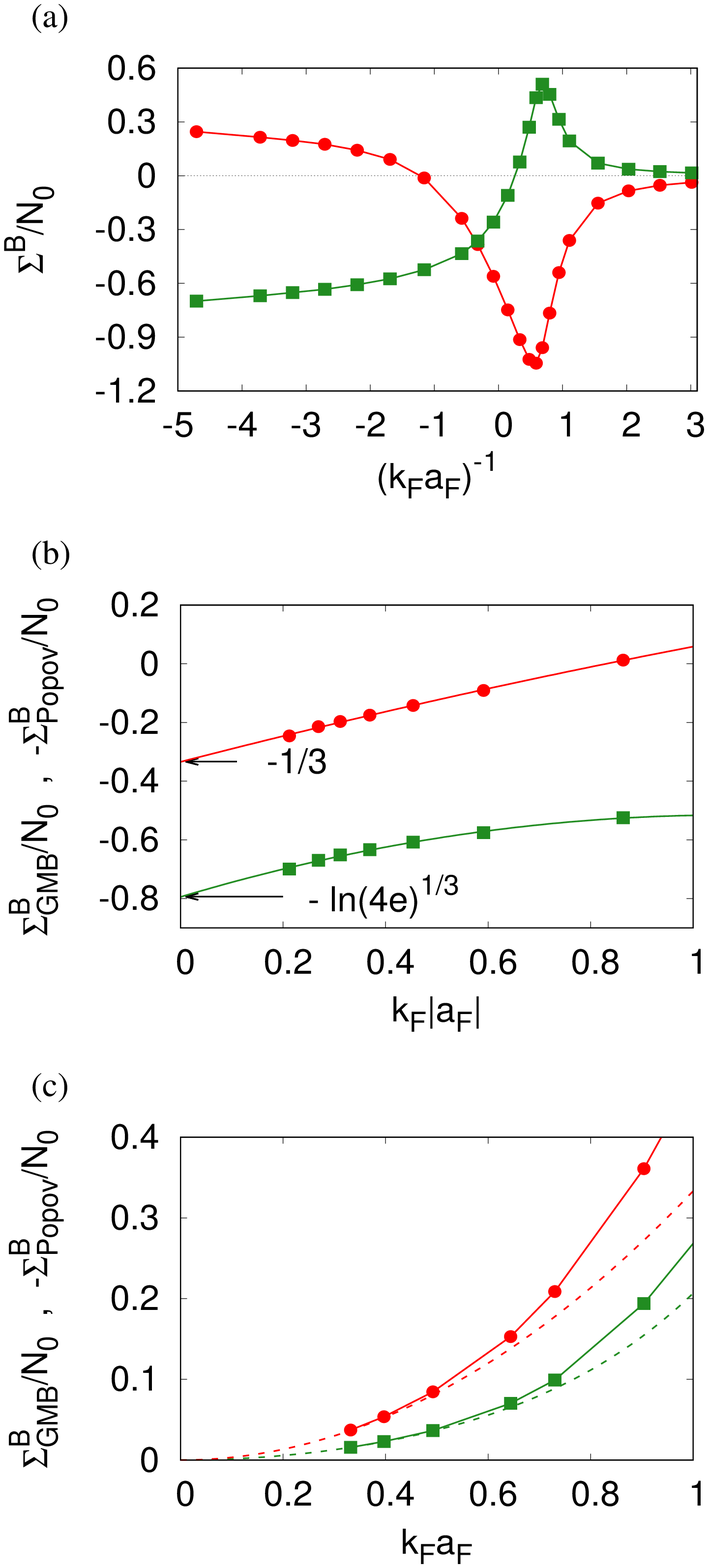}
\caption{(Color online) (a) Bosonic-like self-energies $\Sigma_{\mathrm{Popov}}^{\mathrm{B}}$ (circles) and $\Sigma_{\mathrm{GMB}}^{\mathrm{B}}$ (squares) (in units of the single-particle 
                                          density of states $N_{0}$) vs the coupling $(k_{F}a_{F})^{-1}$. Both quantities are calculated with the values of $(\bar{T}_{c},\bar{\mu}_{c})$ of the full theory.
                                     (b) - $\Sigma_{\mathrm{Popov}}^{\mathrm{B}}$ and $\Sigma_{\mathrm{GMB}}^{\mathrm{B}}$ vs $k_{F}|a_{F}|$ (with $a_{F} < 0$) obtained numerically in the interval $(0,1)$. 
                                           The limiting values for $k_{F}|a_{F}| \rightarrow 0$ are shown to recover the results (\ref{Popov-self-energy-approximate-BCS}) and (\ref{average-particle-hole-bubble}), respectively,
                                           through an extrapolation procedure represented in each case by a full line.
                                      (c) - $\Sigma_{\mathrm{Popov}}^{\mathrm{B}}$ and $\Sigma_{\mathrm{GMB}}^{\mathrm{B}}$ vs $k_{F}a_{F}$ (with $a_{F} > 0$) obtained numerically (symbols plus full lines) in the 
                                            interval $(0,1)$ are compared, respectively, with the analytic behaviours (\ref{Popov-self-energy-approximate-BEC}) and (\ref{GMB-approximation-BEC}) (dashed lines).
                                           [Note that in panels (b) and (c) $\Sigma_{\mathrm{Popov}}^{\mathrm{B}}$ is multiplied by a minus sign.]}
\label{Figure-5}
\end{center} 
\end{figure}

A plot of $\Sigma_{\mathrm{Popov}}^{\mathrm{B}}$ and $\Sigma_{\mathrm{GMB}}^{\mathrm{B}}$ is shown in Fig.~\ref{Figure-5}(a) throughout the BCS-BEC crossover, with both quantities calculated
at the values of $(\bar{T}_{c},\bar{\mu}_{c})$ of the full theory.
Note that these two quantities have somewhat opposite behavior, with a comparable magnitude across the whole crossover.
The limiting behaviors of both quantities in the (extreme) BCS limit are shown in Fig.~\ref{Figure-5}(b).
In particular, the limiting BCS values for $(k_{F}a_{F}) \rightarrow 0^{-}$ have been obtained in both cases by fitting the numerical results with a quadratic polynomial and then extrapolating the curve to $k_{F} a_{F}=0$, recovering in this way with very good accuracy the Popov value $1/3$ given by Eq.~(\ref{Popov-self-energy-approximate-BCS}) and the GMB value $-\ln (4e)^{1/3}$ given by 
Eq.~(\ref{average-particle-hole-bubble}) (in units of $N(\mu)$ with $\mu=E_{F}$).
Figure~\ref{Figure-5}(c) shows further the limiting behaviors of the Popov and GMB bosonic-like self-energies in the BEC limit for small values of $k_{F}a_{F}$.
In this case, our numerical calculations are compared with the analytic expressions for the Popov [Eq.~(\ref{Popov-self-energy-approximate-BEC})] and GMB [Eq.~(\ref{GMB-approximation-BEC})] contributions.
These limiting analytic behaviors are quite well reproduced by our numerical calculation, thus providing a stringent test on its accuracy.

\begin{figure}[h]
\begin{center}
\includegraphics[width=8.0cm,angle=0]{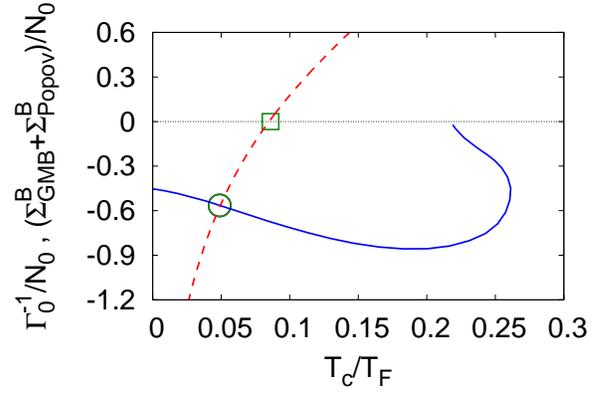}
\caption{(Color online) The graphical procedure for determining numerically the intersection between $\Gamma_{0}^{-1}(Q=0)$ (dashed line) 
                                     and $\Sigma_{\mathrm{GMB}}^{\mathrm{B}} + \Sigma_{\mathrm{Popov}}^{\mathrm{B}}$ (full line)
                                     is shown for the coupling $(k_{F} a_{F})^{-1} = -1.0$ (both quantities are in units of single-particle density of states $N_{0}$). 
                                     Here, $T_{c}^{\mathrm{t}-\!\mathrm{matrix}}=  0.08485 T_{F}$ corresponds to the intersection given by the square and 
                                     $T_{c}^{\mathrm{Popov + GMB}} = 0.04860 T_{F}$ corresponds to the intersection given by the circle.}
\label{Figure-6}
\end{center} 
\end{figure}

Figure~\ref{Figure-6} reports an example about the way the generalized Thouless criterion (\ref{generalized-Thouless-criterion-final-form}) is solved in practice for given coupling.
It amounts to finding the intersection between the curves of $\Gamma_{0}^{-1}(Q=0)$ and $\Sigma_{\mathrm{GMB}}^{\mathrm{B}} + \Sigma_{\mathrm{Popov}}^{\mathrm{B}}$ vs $T/T_{F}$.
One begins by drawing the function $\Gamma_{0}^{-1}(Q=0;T_{c},\mu_{c})$ vs $T_{c}/T_{F}$ (dashed line), where the value of $T_{c}$ and of the associated $\mu_{c}$ are consistent with 
the density equation (\ref{density}).
Since this function given by Eq.~(\ref{Gamma-0-1}) depends explicitly on the coupling, the plot reported in Fig.~\ref{Figure-6} corresponds to a specific value of $(k_{F}a_{F})^{-1}$.
The value of $T_{c}$ at which $\Gamma_{0}^{-1}$ crosses zero (identified by the square in Fig.~\ref{Figure-6}) then corresponds to the (non-self-consistent) t-matrix approximation - 
cf. the discussion of subsection~\ref{sec:numerical-results}-A.
Next one draws the bosonic-like self energy $\Sigma_{\mathrm{GMB}}^{\mathrm{B}} + \Sigma_{\mathrm{Popov}}^{\mathrm{B}}$ vs $T_{c}/T_{F}$ (full line), which also depends on the pair $(T_{c},\mu_{c})$ as specified above.
The intersection of this curve with the function $\Gamma_{0}^{-1}$ (identified by the circle in Fig.~\ref{Figure-6}) provides eventually the value of $T_{c}$ (and thus also of $\mu_{c}$) with the Popov and GMB corrections included
for the given coupling.

\vspace{0.05cm}
\begin{center}
{\bf C. Critical temperature}
\end{center}
\vspace{-0.2cm}

\begin{figure}[t]
\begin{center}
\includegraphics[width=8.5cm,angle=0]{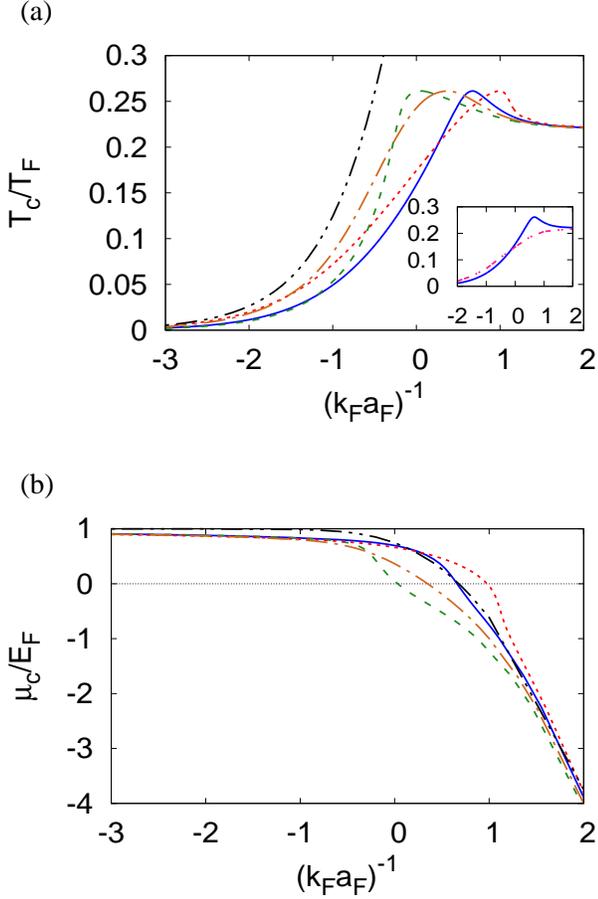}
\caption{(Color online) (a) Five different approximations for the critical  temperature $T_{c}$ are shown vs the coupling $(k_{F}a_{F})^{-1}$: 
                                          $T_{c}^{\mathrm{(GMB+Popov)}}$ (full line);
                                          $T_{c}^{\mathrm{(GMB)}}$ (dashed line);
                                          $T_{c}^{\mathrm{(Popov)}}$ (dotted line);
                                          $T_{c}^{\mathrm{(t-matrix)}}$ (dashed-dotted line);
                                          $\bar{T}_{c}^{(\mathrm{BCS})}$ at the mean-field level (dashed-double-dotted line).
                                          The inset compares the coupling dependence of $T_{c}^{\mathrm{(GMB+Popov)}}$ (full line) and of $T_{c}$ obtained in Refs.~\cite{Haussmann-1994,Haussmann-2007}
                                          within the self-consistent t-matrix approximation (dashed line).
                                     (b) The corresponding values of the chemical potential $\mu_{c}$ evaluated at $T_{c}$ are shown with the same conventions of panel (a).
                                           Both $T_{c}$ and $\mu_{c}$ are in units of the Fermi energy $E_{F}$.}
\label{Figure-7}
\end{center} 
\end{figure}

The complete dependence of the critical temperature $T_{c}$ on coupling obtained in this way is reported in Fig.~\ref{Figure-7}(a).
This figure shows the results obtained by several approximations:
(i) The most complete result $T_{c}^{\mathrm{(GMB+Popov)}}$ obtained by including \emph{both\/} $\Sigma_{\mathrm{GMB}}^{\mathrm{B}}$ and $\Sigma_{\mathrm{Popov}}^{\mathrm{B}}$ (full line);
(ii) The partial result $T_{c}^{\mathrm{(GMB)}}$ obtained by including $\Sigma_{\mathrm{GMB}}^{\mathrm{B}}$ only (dashed line);
(iii) The partial result $T_{c}^{\mathrm{(Popov)}}$ obtained by including $\Sigma_{\mathrm{Popov}}^{\mathrm{B}}$ only (dotted line);
(iv) $T_{c}^{\mathrm{(t-matrix)}}$ corresponding to the (non-self-consistent) t-matrix approximation (dashed-dotted line);
(v) The BCS result $\bar{T}_{c}^{(\mathrm{BCS})}$ obtained at the mean-field level (i.e., with no inclusion of pairing fluctuations) throughout the BCS-BEC crossover \cite{Perali-2004} (dashed-double-dotted line), which extends the expression (\ref{Tc-BCS-no_cutoff}) away from the extreme weak coupling.

From this figure it appears that the curves (i)-(iii) are obtained from the curve (iv) by a non-uniform ``stretching'' of the coupling axis.
Consistently with this observation, all curves (i)-(iv) have a maximum with the same height, although shifted at different couplings.
On the other hand, this maximum is absent when the self-consistent t-matrix approximation of Refs.~\cite{Haussmann-1994,Haussmann-2007} is adopted to calculate $T_{c}$, as shown in the inset of Fig.~\ref{Figure-7}(a) (dashed line) where a comparison with our most complete result $T_{c}^{\mathrm{(GMB+Popov)}}$ is also reported (full line).
Note that these two curves cross each other at about unitarity.

Quite generally, the presence of a maximum in the curve of the critical temperature vs coupling throughout the BCS-BEC crossover would be required by a general argument that, when approaching the extreme BEC limit where composite bosons can be treated as  point-like for all practical purposes, the Bose-Einstein condensation temperature $T_{\mathrm{BEC}}$ for non-interacting bosons should be approached \emph{from above} as shown in Ref.~\cite{Baym-1999}.

Figure~\ref{Figure-7}(b) reports the results for the chemical potential $\mu_{c}$ associated with the values of the critical temperature $T_{c}$ corresponding to the approximations of Fig.~\ref{Figure-7}(a). 
Note, in particular, that the BCS result for $\mu_{c}$ (dashed-double-dotted line) corresponds to the non-interacting value taken at the temperature $T_{c}$.
Note also that our most complete result for $\mu_{c}$, obtained by including \emph{both\/} $\Sigma_{\mathrm{GMB}}^{\mathrm{B}}$ and $\Sigma_{\mathrm{Popov}}^{\mathrm{B}}$ (full line), gives larger values with respect to the (non-self-consistent) t-matrix approximation (dashed-dotted line), while the self-consistent t-matrix approximation of Refs.~\cite{Haussmann-1994,Haussmann-2007} gives a smaller value for 
$\mu_{c}$ at unitarity.
We shall comment more extensively on this issue in Section~\ref{sec:conclusions}.

\begin{figure}[h]
\begin{center}
\includegraphics[width=8.5cm,angle=0]{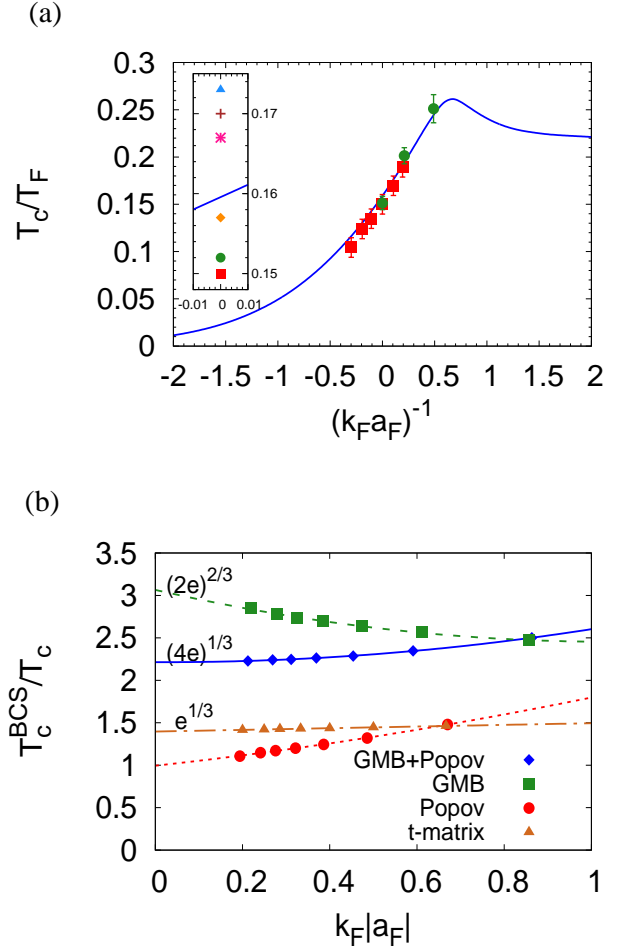}
\caption{(Color online) (a) The results for the critical temperature $T_{c}^{\mathrm{(GMB+Popov)}}$ (full line) are compared with the Quantum Monte Carlo data from Ref.~\cite{Bulgac-2008} (squares with error bars)
                                          and Ref.~\cite{Burovski-2008} (dots with error bars) over an extended region of coupling.
                                          The inset focuses on the results at unitarity, by comparing our value of $T_{c}^{\mathrm{(GMB+Popov)}}$ (full line) with the results from 
                                          Refs.~\cite{Bulgac-2008,Burovski-2008,Nascimbene-2010,Ku-2012,Horikoshi-2010,Goulko-2010} (corresponding to the symbols from bottom to top, in the order).
                                          (b) The ratio $T_{c}^{\mathrm{(BCS)}}/T_{c}$ is shown in the BCS (weak-coupling) regime $k_{F}|a_{F}| \lesssim 1.0$ (with $a_{F}<0$), for the various approximations for $T_{c}$
                                          (GMB+Popov, GMB, Popov, and t-matrix) reported in Fig.~\ref{Figure-7}(a) over a more extended region of coupling. Here, $T_{c}^{\mathrm{(BCS)}}$ is given 
                                          by the expression (\ref{Tc-BCS-no_cutoff}) that holds in the BCS regime.}
\label{Figure-8}
\end{center} 
\end{figure}

Figure~\ref{Figure-8}(a) compares our most complete results (GMB + Popov) for $T_{c}$ over an extended region of the coupling parameter $(k_{F}a_{F})^{-1}$ with the Quantum Monte Carlo (QMC) data available from Refs.~\cite{Bulgac-2008,Burovski-2008}.
The agreement between our results and the QMC data appears quite remarkable, taking into account the fact that our calculations contain no fitting parameters.
It is also worth noting that the value of $T_{c}$ for $(k_{F}a_{F})^{-1}=+0.5$ from Ref.~\cite{Burovski-2008} is larger than the BEC value $T_{\mathrm{BEC}}$ attained by our calculation in the extreme BEC limit, thus supporting the  presence of a maximum in the curve of $T_{c}$ vs $(k_{F}a_{F})^{-1}$.
In addition, the inset of Fig.~\ref{Figure-8}(a) compares our result for $T_{c}$ at unitarity with the corresponding data reported both in experimental \cite{Nascimbene-2010,Ku-2012,Horikoshi-2010} and theoretical (QMC) \cite{Bulgac-2008,Burovski-2008,Goulko-2010} works.
Also in this case, it is remarkable that our value for $T_{c}$ lies well within the boundaries provided by these data.
 
Finally, Fig.~\ref{Figure-8}(b) considers the ratio of the BCS critical temperature $T_{c}^{\mathrm{(BCS)}}$ given by Eq.~(\ref{Tc-BCS-no_cutoff}), which well approximates the mean-field result for $T_{c}$ in the BCS (weak-coupling) 
regime $k_{F}|a_{F}| \lesssim 1.0$ with $a_{F}<0$, with our numerical results for $T_{c}$ that were reported in Fig.~\ref{Figure-7}(a) over a more extended region of coupling.
The numerical results (symbols) have been extrapolated to the limit $k_{F}|a_{F}| \rightarrow 0$ through the fittings curves also reported in the figure (lines), where the limiting values coincide in each case  with those obtained analytically in subsections~\ref{sec:G-MB-BCS-BEC}-C and \ref{sec:G-MB-BCS-BEC}-E.
This represents a further strong check on the accuracy of our calculations.

\vspace{0.05cm}
\begin{center}
{\bf D. Relevance of the dependence on wave-vector and frequency of the pair propagators entering the GMB bosonic self-energy}
\end{center}
\vspace{-0.2cm}

We have emphasized in the Introduction that this paper includes for the first time the full wave-vector and frequency dependence of the pair propagator $\Gamma_{0}$ in the calculation of the GMB contribution.
We have also shown throughout this work that taking into account this dependence is essential for a correct calculation of the GMB contribution away from the (extreme) BCS regime.
In this context, the question naturally arises about which one of these two dependences (that is, on wave vector or frequency) is the dominant one.

\begin{figure}[t]
\begin{center}
\includegraphics[width=8.5cm,angle=0]{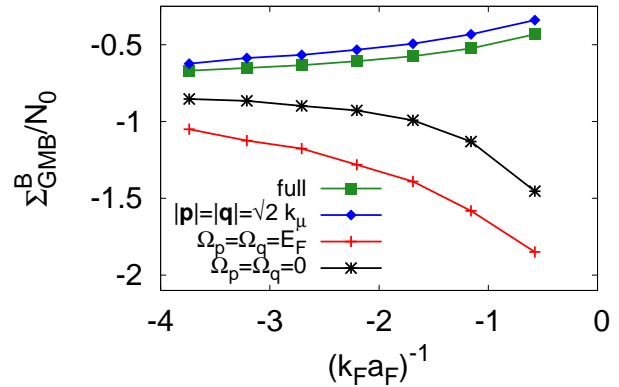}
\caption{(Color online) The full calculation of $\Sigma_{\mathrm{GMB}}^{\mathrm{B}}$ vs $(k_{F}a_{F})^{-1}$ is compared with partial calculations of the same quantity, where either the wave-vector 
                                    or the frequency dependence has been neglected in both pair propagators $\Gamma_{0}$ entering the expression of $\Sigma_{\mathrm{GMB}}^{\mathrm{B}}$. 
                                    The pairs $(\bar{T}_{c},\bar{\mu}_{c})$ obtained by the full calculation are also used in the partial calculations.}
\label{Figure-9}
\end{center} 
\end{figure}

To answer this question, we have performed additional ``partial calculations'' of the GMB bosonic-like self-energy $\Sigma_{\mathrm{GMB}}^{\mathrm{B}}$, where the dependence on \emph{either\/} the wave vector \emph{or\/} the frequency has been neglected in both pair propagators $\Gamma_{0}$ that appear therein (while using the same values of $(\bar{T}_{c},\bar{\mu}_{c})$ in all calculations).
Specifically, in each $\Gamma_{0}$ we have: (i) Either set the magnitude of the wave vectors $|\mathbf{p}|$ and $|\mathbf{q}|$ equal to $\sqrt{2}k_{\mu}$ (where the absolute value of the corresponding integrand is maximum) while maintaining the full frequency dependence; (ii) Or set the frequency alternatively equal to $0$ or to $E_{F}$ while maintaining the full wave-vector dependence.
The results of these calculations are reported in Fig.~\ref{Figure-9} on the BCS side of unitarity, where they are compared with the full calculation with both the wave-vector and frequency dependences of $\Gamma_{0}$ taken into account.
From this comparison one concludes that in the calculation of $\Sigma_{\mathrm{GMB}}^{\mathrm{B}}$, not only the frequency dependence plays a dominant role over the wave-vector dependence of $\Gamma_{0}$, but also that neglecting the frequency dependence of $\Gamma_{0}$ yields a result considerably different from that of the full calculation, even with the wrong curvature of $\Sigma_{\mathrm{GMB}}^{\mathrm{B}}$ as a function of coupling.

\section{Concluding remarks and perspectives}
\label{sec:conclusions}

In this paper, we have dealt with the calculation of the GMB correction in such a way that we could consistently extend it to scan the whole BCS-BEC crossover, whereby largely overlapping Cooper pairs on the BCS side evolve continuously into dilute composite bosons on the BEC side.
The GMB correction was, in fact, originally introduced \cite{GMB-1961} to complement the BCS theory of superconductivity \cite{BCS-1957}, which at that time was meant to apply only to largely overlapping Cooper pairs.
For this reason, the GMB theory took advantage of specific approximations which are valid under these circumstances, resulting into a sizable reduction (by a factor of $2.2$) of the value of the critical temperature with respect to the standard BCS value.
The recent advent of accurate experiments with ultra-cold Fermi gases has then brought up the need for an accurate calculation of the GMB correction throughout the BCS-BEC crossover, thereby avoiding those approximations that would apply specifically to the BCS limit but could not be extended to the whole crossover.
Previous attempts by diagrammatic methods to extend the calculation of the GMB correction throughout the BCS-BEC crossover \cite{Yu-2009,Ruan-2013} apparently did not realize this delicate point and made an incorrect use of the same main approximations utilized in the original GMB paper.

Our handling of the GMB correction throughout the BCS-BEC crossover has helped clarifying an important physical point, that is, that the effective disentanglement between the particle-particle excitations (which are characteristic of superconductivity) and the particle-hole excitations (which are characteristic of screening) holds only in the (extreme) BCS limit of the crossover, which was of exclusive interest to the original GMB approach \cite{GMB-1961}.
We have shown that, consistently with the key ingredients on which any physical sensible theoretical treatment of the BCS-BEC crossover must rely on, the above disentanglement between particle-particle and particle-hole excitations is not bound to occur when moving away from the BCS limit.
In this context, we have also shown [cf. Appendix~\ref{sec:appendix-B}] that the GMB correction, apart from maintaining its role in the BCS limit, acquires also an important role in the BEC limit of the crossover where it contributes significantly to the value of the scattering length $a_{B}$ of composite bosons.
Otherwise, if one would (incorrectly) stick to maintain the disentanglement between particle-particle and particle-hole excitations even in the BEC limit, the GMB correction would become totally irrelevant in this limit.

From the computational side, we have performed a very accurate numerical calculation of the GMB correction, maintaining the full dependence on wave vector and frequency of the pair propagators that appear in its expression. 
In this respect, we have been able to reproduce in a \emph{totally} numerical fashion the factor of $2.2$ for the reduction of the critical temperature with respect to the standard BCS value (while originally this result was obtained in an analytic way \cite{GMB-1961}).
The accuracy of our numerical calculations was further tested against the analytic results that can be obtained both in the BCS and BEC limits of the crossover.
To this end, we found it necessary to complement the GMB correction by a further correction based on the Popov theory for point-like bosons, which provides an important mean-field shift contribution to the chemical potential of the constituent fermions in the BCS limit and to the chemical potential of the composite bosons in the BEC limit.

A definite success of our accurate simultaneous numerical handling of the GMB and Popov corrections throughout the BCS-BEC crossover is represented by the remarkable agreement we have obtained, between our calculated values of the critical temperature and those available by QMC calculations in the core of the crossover region (whereby $-0.5 \lesssim (k_{F} a_{F})^{-1} \lesssim +0.5$) and by experiments with ultra-cold Fermi gases at unitarity.
This agreement appears particularly significative, in the light of the fact that our first-principles calculations do not contain any fitting parameter. 

In this context, however, it should be pointed out that our calculation does not match the value of the chemical potential $\mu_{c}$ at $T_{c}$ and unitarity, within the range determined by an available experiment and by alternative theoretical calculations.
Specifically, at unitarity and at the respective value of $T_{c}$, $\mu_{c}/E_{F}$ equals $0.3659$ within the non-self-consistent t-matrix approximation, $0.6971$ by adding to it the GMB and Popov corrections as it was done in the present work, $0.394$ within the self-consistent t-matrix approximation of Refs.~\cite{Haussmann-1994,Haussmann-2007}, and $0.42$ as obtained experimentally in Ref.~\cite{Ku-2012}.
It thus appears that, adding the GMB and Popov corrections on top of the non-self-consistent t-matrix approximation, makes the agreement with the experimental chemical potential worse than that obtained with the non-self-consistent t-matrix approximation itself.

This failure in obtaining a reasonable value of the thermodynamic chemical potential can be attributed to the lack of (at least some degree of) self-consistency in the fermionic propagators $G_{0}$ entering the diagrammatic structures of $\Sigma_{\mathrm{Popov}}^{\mathrm{B}}$ [cf. Eq.~(\ref{Popov-self-energy-definition})] and $\Sigma_{\mathrm{GMB}}^{\mathrm{B}}$ [cf. Eq.~(\ref{GMB-self-energy-definition})], as well as of $\Gamma_{0}$ [cf. Eqs.~(\ref{Gamma-0}) and (\ref{particle-particle-bubble})] and of the fermionic self-energy $\Sigma$ [cf. Eq.~(\ref{Sigma-NSR})].
To estimate the effect of introducing this self-consistency on the critical temperature obtained by including the Popov and GMB corrections, we can parallel the way the self-consistency was implemented in this paper for the bosonic-like pair propagator $\Gamma_{0}$.
To this end, self-consistency in the fermionic propagators can approximately be dealt with at any given coupling by including a suitable \emph{constant} (that is, independent of the four-vector $k$) self-energy shift 
$\Sigma_{0}$ in each of the above propagators $G_{0}$.
Accordingly, this constant shift $\Sigma_{0}$ would get subtracted from the chemical potential $\mu$, such that $\mu \rightarrow \mu - \Sigma_{0} = \mu'$.
The only way this replacement would not affect the values of the critical temperature, as determined above within the full theory that includes $\Sigma_{\mathrm{Popov}}^{\mathrm{B}}$ and 
$\Sigma_{\mathrm{GMB}}^{\mathrm{B}}$ [cf. the full line in Fig.~\ref{Figure-7}(a)], would be that $\mu'$ thus determined corresponds to the chemical potential reported in Fig.~\ref{Figure-7}(b) (full line).
\begin{figure}[t]
\begin{center}
\includegraphics[width=8.7cm,angle=0]{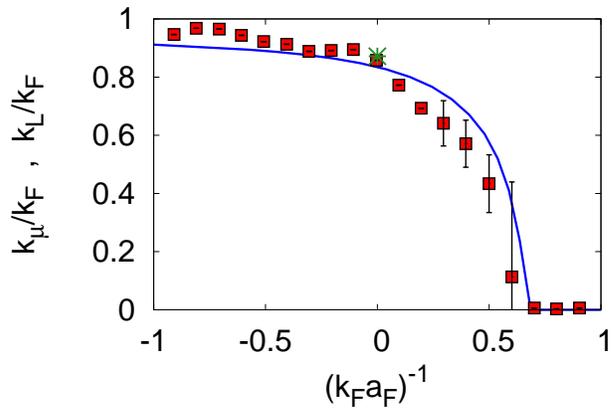}
\caption{(Color online) The Luttinger wave vector $k_{L}$, obtained in Refs.~\cite{Perali-2011,Palestini-2013} from the single-particle spectral function to identify the existence of an underlying 
                                    Fermi surface in a Fermi gas with attractive inter-particle interaction, is shown at $T_{c}$ vs coupling across the BCS-BEC crossover (squares with error bars).
                                    The corresponding value obtained at unitarity from the QMC calculation of Ref.~\cite{Carlson-2005} is also reported for comparison (star).
                                    This quantity is compared with the values of $k_{\mu}=\sqrt{2m \mu}$ for $\mu>0$ (full curve) obtained from the values of $\mu$ corresponding to the full curve of Fig.~\ref{Figure-7}(b).}
\label{Figure-10}
\end{center} 
\end{figure}
An important observation can be made at this point, that the same value of $\mu'$ also identifies an underlying Fermi surface with radius $k_{\mu'} = \sqrt{2 m \mu'}$ when $\mu' > 0$, at whose location a
back-bending occurs in the dispersion relation $\sqrt{\left(\frac{\mathbf{k}^{2}}{2m} - \mu' \right)^{2} + \tilde{\Delta}^{2}}$ as obtained from the single-particle spectral function, where $\tilde{\Delta}$ can be identified with a pairing gap in the superfluid phase \cite{Pisani-2004} or with a pseudo-gap in the normal phase \cite{Perali-2011}.
Figure~\ref{Figure-10} shows the value of $k_{\mu'}$ obtained in Refs.~\cite{Perali-2011,Palestini-2013} (where it was referred to as the Luttinger wave vector $k_{L}$) from a study of the single-particle spectral function at $T_{c}$ vs coupling (squares with error bars).
These data are compared with the value of $k_{\mu}=\sqrt{2m \mu}$ for $\mu>0$ (full curve) obtained from the corresponding full curve of Fig.~\ref{Figure-7}(b).
We recall that the values of $k_{L}$ reported in Fig.~\ref{Figure-10} have been validated by an extensive comparison with experiment over an extended range of coupling about unitarity \cite{Gaebler-2010,Perali-2011} and also by the value obtained at unitarity by the QMC calculation of Ref.~\cite{Carlson-2005} (identified by the star in Fig.~\ref{Figure-10}).
The quite good overall agreement obtained in this figure confirms our identification between $\mu$ of the present theory (the full curve of Fig.~\ref{Figure-7}(b)) with $\mu' = \mu - \Sigma_{0}$ of a more refined theory that would include fermionic self-consistency, and supports our argument that the values of $T_{c}$ obtained in this paper by including the Popov and GMB corrections [the full line in Fig.~\ref{Figure-7}(a)] should not be affected by including explicitly this self-consistency.

A future natural extension of the present approach will be to consider the superfluid phase below $T_{c}$ and calculate the pairing gap $\Delta$ down to zero temperature.
Actually, this problem was considered in the original GMB paper \cite{GMB-1961}, where it was found that at zero temperature also the value of $\Delta$ is reduced by a factor $2.2$ with respect to the BCS result.
The challenge now would be to extend this GMB result to the whole BCS-BEC crossover for all temperatures between zero and $T_{c}$.
To this end, a suitable diagrammatic form of the gap equation should be set up beforehand, which would allow us to take into account the GMB as well as the Popov correction extended to the superfluid 
phase.
This topic will be postponed to future work.

Extending the proper treatment of the GMB correction to the superfluid phase across the BCS-BEC crossover appears particularly relevant at this time, since accurate experimental values of the pairing gap were recently made available via Bragg spectroscopy with ultra-cold Fermi gases \cite{Vale-2017}.
This topic is also of much interest in the context of nuclear physics, where the value of the pairing gap was calculated at zero temperature by QMC methods essentially over the whole BCS side of unitarity 
\cite{Carlson-2010}.


\begin{center}
\begin{small}
{\bf ACKNOWLEDGMENTS}
\end{small}
\end{center}
\vspace{-0.2cm}

We are indebted to S. Girotti for help during the initial stage of this work.
This work was partially supported by the Italian MIUR under Contract PRIN-2015 No. 2015C5SEJJ001.

\appendix                                                                                                                                                                                                                                                                                                                                                                                                        
\section{IMPLEMENTING THE NUMERICAL CALCULATION OF THE POPOV AND GMB DIAGRAMS}
\label{sec:appendix-A}

We pass now to discuss in detail the procedures that we have adopted for implementing the numerical calculations of the Popov [Eq.~(\ref{Popov-self-energy-definition})] and GMB 
[Eq.~(\ref{GMB-self-energy-definition})] diagrammatic contributions (both with $Q=0$).
A detailed discussion is relevant, in the light of the fact that in these calculations we have included the \emph{full}  wave-vector and frequency dependence of the pair propagators
$\Gamma_{0}$ appearing in the expressions (\ref{Popov-self-energy-definition}) and (\ref{GMB-self-energy-definition}).
In particular, this is especially important for the GMB expression (\ref{GMB-self-energy-definition}), for which this inclusion was never attempted before.
In the following, we discuss the two contributions separately.

For the sake of definiteness, the procedures to calculate numerically the integrals and sums entering the Popov and GMB diagrammatic contributions will be discussed below in terms of the ``bare'' pair propagator $\Gamma_{0}$.
In practice, however, these calculations will be performed utilising instead the ``dressed'' pair propagator $\Gamma$ of Eq.~(\ref{Gamma-1}) in the place of $\Gamma_{0}$.
This replacement will obviously not affect the procedures described below, which rest on taking full account of the wave-vector and frequency dependence of $\Gamma_{0}$ (or, equivalently,
of $\Gamma$).

\vspace{0.05cm}
\begin{center}
{\bf Popov contribution}
\end{center}
\vspace{-0.2cm}

For the sake of the following discussion, it is convenient to reproduce here the expression (\ref{Popov-self-energy-definition}) with $Q=0$:
\begin{eqnarray}
\Sigma_{\mathrm{Popov}}^{\mathrm{B}} & = & - 2  \int \! \frac{d\mathbf{k}}{(2 \pi)^3} \, T \, \sum_{n} \int \! \frac{d\mathbf{q}}{(2 \pi)^3} \, T \sum_{\nu_{q}} 
\nonumber \\
& \times & G_{0}(k)^{2} \, G_{0}(-k) \, G_{0}(q-k) \, \Gamma_{0}(q) 
\label{Popov-self-energy-definition-Q=0}
\end{eqnarray}
\noindent
with the four-vector notation $k=(\mathbf{k},\omega_{n})$ and $q=(\mathbf{q},\Omega_{\nu_{q}})$, where $\omega_{n}$ and $\Omega_{\nu_{q}}$ are fermionic and bosonic Matsubara frequencies, respectively.
Here, $G_{0}(k) = (i\omega_{n} - \xi_{\mathbf{k}})^{-1}$ with $\xi_{\mathbf{k}}=\mathbf{k}^2/(2m) - \mu$ and $\Gamma_{0}$ is given by Eq.~(\ref{Gamma-0}). 
The expression (\ref{Popov-self-energy-definition-Q=0}) contains summations over two Matsubara frequencies, and integrations over four angles (three of which turn out to be trivial) and over two magnitudes of wave vectors.
We have optimized these sums and integrations, by performing them in the following order.

\vspace{0.1cm}
\noindent
{\bf \emph{Sum over the fermionic Matsubara frequency:}}
The sum over the fermionic Matsubara frequency $\omega_{n}$ can be done analytically.
One arrives at the following expression:
\begin{equation}
\Sigma_{\mathrm{Popov}}^{\mathrm{B}} = - 2 \!\! \int \!\! \frac{d\mathbf{k}}{(2 \pi)^3} \! \int \!\! \frac{d\mathbf{q}}{(2 \pi)^3} \, T \sum_{\nu_{q}} \, I(\xi_{\mathbf{k}},\xi_{\mathbf{q-k}};\Omega_{\nu_{p}},\Omega_{\nu_{q}}) \, \Gamma_{0}(q) 
\label{Popov-self-energy-Q=0-first-sum-done}
\end{equation}
\noindent
where
\begin{widetext}
\begin{eqnarray}
I(\xi_{\mathbf{k}},\xi_{\mathbf{q-k}};\Omega_{\nu_{p}},\Omega_{\nu_{q}}) & = & - \frac{f(\xi_{\mathbf{k}})}{ (2 \xi_{\mathbf{k}})^{2} (\xi_{\mathbf{k}}+\xi_{\mathbf{q-k}}-i \Omega_{\nu_{q}}) } 
 -  \frac{f(\xi_{\mathbf{k}})}{ 2 \xi_{\mathbf{k}} (\xi_{\mathbf{k}}+\xi_{\mathbf{q-k}}-i \Omega_{\nu_{q}})^{2} }
   +  \frac{df(\xi_{\mathbf{k}})/d\xi_{\mathbf{k}}}{ 2 \xi_{\mathbf{k}} (\xi_{\mathbf{k}}+\xi_{\mathbf{q-k}}-i \Omega_{\nu_{q}})}
\nonumber \\  
& + & \frac{f(-\xi_{\mathbf{k}})}{ (2 \xi_{\mathbf{k}})^{2} (-\xi_{\mathbf{k}}+\xi_{\mathbf{q-k}}-i \Omega_{\nu_{q}}) }  
- \frac{f(-\xi_{\mathbf{q-k}})}{ (\xi_{\mathbf{k}}+\xi_{\mathbf{q-k}}-i \Omega_{\nu_{q}})^{2} (-\xi_{\mathbf{k}}+\xi_{\mathbf{q-k}}-i \Omega_{\nu_{q}}) } \, .
\label{function-I} 
\end{eqnarray}
\end{widetext}

\vspace{0.1cm}
\noindent
{\bf \emph{Angular integrations over the wave vectors:}}
The expression (\ref{function-I}) depends only on the relative angle between $\mathbf{k}$ and $\mathbf{q}$.
The integration over this angle is performed numerically with $100$ points, while the integrations over the remaining three angles (which do not appear explicitly in the expression (\ref{function-I})) contribute a mere numerical factor $8 \pi^{2}$.

\vspace{0.1cm}
\noindent
{\bf \emph{Radial integration over the fermionic wave vector:}}
The radial integration over the magnitude $|\mathbf{k}|$ of the fermionic wave vector $\mathbf{k}$ is conveniently split into three intervals, namely, $[0,k_{c}]$, 
$[k_{c},k_{c}+2k_{F}]$, and $[k_{c}+2k_{F},+\infty]$, where
\begin{equation}
k_{c} = \sqrt{2 m \left(\mu^{2}+T^{2}\right)^{1/2}}
\label{k-c}
\end{equation}
\noindent
irrespective of the sign of $\mu$. 
In each interval, $50$ integration points at most prove sufficient. 
The need for introducing the wave vector $k_{c}$ stems from the need for reproducing with good accuracy the shape of the peak developed by the integrand at about $k_{c}$.
The second cutoff at $k_{c}+2k_{F}$ is instead required to deal with the large-$|\mathbf{k}|$ tail of the integrand. 
[Similar considerations (about the peaks and tails of the integrands) will apply when introducing the cutoffs $\tilde{q}_{c}$, $(k_{c_{1}},k_{c_{2}})$, $(q_{c_{1}},q_{c_{2}})$, and $(p_{c_{1}},p_{c_{2}})$ later on in this Appendix.]

\vspace{0.1cm}
\noindent
{\bf \emph{Sum over the bosonic Matsubara frequency:}}
It turns out that the last sum over the Matsubara frequency $\Omega_{\nu_{q}}$ decays like $|\Omega_{\nu_{q}}|^{-1.5}$ for large $|\Omega_{\nu_{q}}|$.
This rather slow decay forces us to perform the sum with special care, by 
(i) summing up the discrete values from $\nu = 0$ up to $\nu_{c_{1}}=500$,
(ii) transforming the discrete sum into an integral from $\Omega_{\nu_{c_{1}}}$ up to $\Omega_{\nu_{c_{2}}}=1.5 \times 10^{6} E_{F}$, and
(iii) calculating the integral analytically from $\nu_{c_{2}}$ up to infinity by estimating the coefficient of the $|\Omega_{\nu_{q}}|^{-1.5}$ power-law decay.
[The sum for $\Omega_{\nu_{q}}<0$ can simply be obtained by complex conjugation.]

\vspace{0.1cm}
\noindent
{\bf \emph{Radial integration over the bosonic wave vector:}}
The last integral over $|\mathbf{q}|$ is divided in three intervals, namely, $[0,\tilde{q}_{c}]$, $[\tilde{q}_{c},3 \tilde{q}_{c}]$, and $[3 \tilde{q}_{c},+\infty]$
where $\tilde{q}_{c}=2 k_{c}$ with $k_{c}$ defined above in Eq.~(\ref{k-c}).
The integration over the first two intervals is performed numerically with $10$ points in each interval, while the integration over the last interval is performed analytically 
by estimating the coefficient of the $|\mathbf{q}|^{-4}$ power-law decay.

\vspace{0.05cm}
\begin{center}
{\bf GMB contribution}
\end{center}
\vspace{-0.2cm}

It is again convenient to reproduce here the expression (\ref{GMB-self-energy-definition}) with $Q=0$:
\begin{widetext}
\begin{equation}
\Sigma_{\mathrm{GMB}}^{\mathrm{B}} = \!\! \int \! \frac{d\mathbf{k}}{(2 \pi)^3} \, T \, \sum_{n} \int \! \frac{d\mathbf{p}}{(2 \pi)^3} \, T \sum_{\nu_{p}} \,
                                                                                                                                                    \int \! \frac{d\mathbf{q}}{(2 \pi)^3} \, T \sum_{\nu_{q}} 
G_{0}(p-k) \, G_{0}(k-p) \, G_{0}(p+q-k) \, G_{0}(k) \, G_{0}(q-k) \, G_{0}(k-q) \, \Gamma_{0}(p) \, \Gamma_{0}(q) 
\label{GMB-self-energy-definition-Q=0}
\end{equation}
\end{widetext}
\noindent
with the additional four-vector notation $p=(\mathbf{p},\Omega_{\nu_{p}})$ with respect to Eq.~(\ref{Popov-self-energy-definition-Q=0}), where $\Omega_{\nu_{p}}$ is a bosonic Matsubara frequency.
The expression (\ref{GMB-self-energy-definition-Q=0}) contains summations over three Matsubara frequencies, and integrations over six angles (three of which turn out to be trivial) and over three magnitudes of wave vectors.
We have optimized these sums and integrations, by performing them in the following order.

\vspace{0.1cm}
\noindent
{\bf \emph{Sum over the fermionic Matsubara frequency:}}
It is convenient to perform the sum over the fermionic Matsubara frequency $\omega_{n}$ first.
This can be done analytically in closed form, owing to the simple expression $G_{0}(k) = (i\omega_{n} - \xi_{\mathbf{k}})^{-1}$ of the \emph{bare} fermionic single-particle propagator.
Otherwise, if one would dress the bare propagator $G_{0}$ with a fermionic self-energy $\Sigma$ (like in Eq.~(\ref{density}), or even in a more complicated fully self-consistent fashion), performing analytically the sum over $\omega_{n}$ would no longer be possible and one should unavoidably revert to a fully numerical evaluation of this sum.

In this way, one arrives at the compact expression:
\begin{widetext}
\begin{equation}
\Sigma_{\mathrm{GMB}}^{\mathrm{B}} = \int \! \frac{d\mathbf{k}}{(2 \pi)^3} \, \int \! \frac{d\mathbf{p}}{(2 \pi)^3} \, T \sum_{\nu_{p}} \, \int \! \frac{d\mathbf{q}}{(2 \pi)^3} \, T \sum_{\nu_{q}} 
\, J(\xi_{\mathbf{k}},\xi_{\mathbf{p-k}},\xi_{\mathbf{q-k}},\xi_{\mathbf{p+q-k}};\Omega_{\nu_{p}},\Omega_{\nu_{q}}) \,  \Gamma_{0}(p) \, \Gamma_{0}(q)
\label{GMB-self-energy-Q=0-first-sum-done}
\end{equation}
\noindent
where 
\begin{eqnarray}
& & J(\xi_{\mathbf{k}},\xi_{\mathbf{p-k}},\xi_{\mathbf{q-k}},\xi_{\mathbf{p+q-k}};\Omega_{\nu_{p}},\Omega_{\nu_{q}}) = 
                             \frac{1}{2} \, \frac{1}{ [\xi_{\mathbf{k}}+\xi_{\mathbf{p+q-k}}-i(\Omega_{\nu_{p}}+ \Omega_{\nu_{q}})] } 
\nonumber \\
& \times &  \left\{ - \frac{ \mathcal{F} ( \xi_\mathbf{p-k} , \xi_{\mathbf{k}} , \Omega_{\nu_{p}}) }{ \xi_{(-)} \xi_{(+)} } 
                           + \frac{ \mathcal{F} ( \xi_\mathbf{q-k} , \xi_{\mathbf{k}} , \Omega_{\nu_{q}}) }{ \xi_{(-)} \xi_{(+)}^{*} }  
                           - \frac{ \mathcal{F} ( -\xi_\mathbf{p-k} , \xi_{\mathbf{k}} , \Omega_{\nu_{p}}) }{ \xi_{(-)}^{*} \xi_{(+)}^{*} } 
                           + \frac{ \mathcal{F} ( -\xi_\mathbf{q-k} , \xi_{\mathbf{k}} , \Omega_{\nu_{q}}) }{ \xi_{(-)}^{*} \xi_{(+)} } 
                           + \frac{ \mathcal{F} ( \xi_\mathbf{p-k} , -\xi_{\mathbf{p+q-k}} , -\Omega_{\nu_{q}}) }{ \xi_{(-)} \xi_{(+)} }   \right.                    
\nonumber \\  
& &         \left.      - \frac{ \mathcal{F} ( \xi_\mathbf{q-k} , -\xi_{\mathbf{p+q-k}} , -\Omega_{\nu_{p}}) }{ \xi_{(-)} \xi_{(+)}^{*} } 
                           + \frac{ \mathcal{F} (- \xi_\mathbf{p-k} , -\xi_{\mathbf{p+q-k}} , -\Omega_{\nu_{q}}) }{ \xi_{(-)}^{*} \xi_{(+)}^{*} } 
                            - \frac{ \mathcal{F} (-  \xi_\mathbf{q-k} , - \xi_{\mathbf{p+q-k}} , -\Omega_{\nu_{p}}) }{ \xi_{(-)}^{*} \xi_{(+)} }  \right\} 
\label{function-J}
\end{eqnarray}
\end{widetext}
\noindent
and
\begin{eqnarray}
\xi_{(\pm)} & = & \xi_\mathbf{p-k} \pm \xi_\mathbf{q-k} + i(\Omega_{\nu_{p}}-\Omega_{\nu_{q}})
\nonumber \\
\mathcal{F}(x,y,z) & = & \frac{f(x) - f(y)}{x(x-y+iz)} \,.
\label{definitions}
\end{eqnarray}

Depending on the values of $\Omega_{\nu_{p}}$ and $\Omega_{\nu_{q}}$, the expression (\ref{function-J}) can be further simplified according to the following steps:
(i) When either $\Omega_{\nu_{p}} \ne 0$ or $\Omega_{\nu_{q}} \ne 0$, the eight terms therein can be reduced to four by  the change of variable $\mathbf{k'} = \mathbf{p+q-k}$ in half of the original terms; 
(ii) When $\Omega_{\nu_{p}} \ne \Omega_{\nu_{q}}$, a further change of variables $p'=q$ and $q'=p$ in half of the four terms that are left after step (i) reduces them to two terms only;
(iii) When $\Omega_{\nu_{p}} = \Omega_{\nu_{q}} \ne 0$, one has instead to stick with the four terms obtained in step (ii);
(iv) Finally, when $\Omega_{\nu_{p}} = \Omega_{\nu_{q}} = 0$ one has to stick with the original eight terms of the expression (\ref{function-J}), in order to avoid introducing unnecessary principal values integrals in the numerical calculation.

\vspace{0.1cm}
\noindent
{\bf \emph{Angular integrations over the wave vectors:}}
The expressions (\ref{function-J}) and (\ref{definitions}) depend explicitly on three angles only.
If one takes the $z$-axis in wave-vector space oriented along the direction of $\mathbf{k}$ and the $x$-axis such that $\mathbf{q}$ belongs to the $x$-$z$ plane, these three angles are the
polar angle $\theta_{\mathbf{q-k}}$ of $\mathbf{q}$ and the azimuthal $\varphi_{\mathbf{p}}$ and polar $\theta_{\mathbf{p-k}}$ angles of $\mathbf{p}$. 
While the integration over $\varphi_{\mathbf{p}}$ can be done analytically, the integrations over $\theta_{\mathbf{p-k}}$ and $\theta_{\mathbf{q-k}}$ have to be performed numerically.
For both integrations $30$ points prove usually sufficient.
Finally, the integrations over the remaining three angles of $(\mathbf{k},\mathbf{p},\mathbf{q})$ (i.e., those on which the expressions (\ref{function-J}) and (\ref{definitions}) do not depend) contribute a numerical factor $8 \pi^{2}$.

\vspace{0.1cm}
\noindent
{\bf \emph{Radial integration over the fermionic wave vector:}}
The radial integration over the magnitude $|\mathbf{k}|$ of the fermionic wave vector $\mathbf{k}$ is conveniently split into three intervals, namely, $[0,k_{c_{1}}]$, $[k_{c_{1}},k_{c_{2}}]$, 
and $[k_{c_{2}},+\infty]$.
Here, $k_{c_{1}}=\mathrm{max}\left\{k_{c},(|\mathbf{p}|+|\mathbf{q}|)/2\right\}$ with $k_{c}$ given by Eq.~(\ref{k-c}), while $k_{c_{2}}=2 k_{c_{1}}$ for $\mu>0$ and $k_{c_{2}}=4 k_{c_{1}}$ 
for $\mu<0$.
In each interval, $15$ integration points prove sufficient.

\vspace{0.1cm}
\noindent
{\bf \emph{Sum over the bosonic Matsubara frequencies:}}	
At this point, it is sufficient to calculate the sums over the bosonic Matsubara frequencies $(\Omega_{p},\Omega_{q})$ in the half-plane $\Omega_{p} \ge 0$ only, since in the other half-plane $\Omega_{p} < 0$ 
the integrand can be obtained by complex conjugation.
In this case, a natural cutoff is given by the frequency $\Omega_{c} = 2 \pi \nu_{c} T = (|\mathbf{p}|^{2} + |\mathbf{q}|^{2})/(2m)$.
The discrete sum over the frequencies $(\Omega_{p},\Omega_{q})$ is then computed over the trapezoidal area $0 \le \Omega_{p} \le \Omega_{c}$ and 
$-(\Omega_{c} + \Omega_{p}) \le \Omega_{q} \le \Omega_{c} + \Omega_{p}$, while outside this area a continuum approximation is adopted which transforms the discrete sums into a two-dimensional integral.
Care should be exerted along the line $\Omega_{p} = \Omega_{q}$ where the integrand presents pronounced peaks.
In the asymptotic regions $\Omega_{p} \gg \Omega_{c}$ and $|\Omega_{q}| \gg \Omega_{c}$ the integrand behaves like $\Omega_{p}^{-3.5}$ and $|\Omega_{q}|^{-3.5}$, such that
$20$ integration points prove sufficient for each variable.

\vspace{0.1cm}
\noindent
{\bf \emph{Radial integrations over the bosonic wave vectors:}}
First, the integral over $|\mathbf{q}|$ is divided in two intervals, namely, $[0,q_{c_{1}}]$ and $[q_{c_{1}},q_{c_{2}}]$, where $q_{c_{1}}=\mathrm{max}\left\{q_{c},|\mathbf{p}|\right\}$ and
$q_{c_{2}}=7k_{F}+|\mathbf{p}|$ with $q_{c}=\sqrt{2} k_{c}$ and $k_{c}$ given by Eq.~(\ref{k-c}).
The integrand turns out to have a minimum at about $q_{c_{1}}$, past which it decays to zero like $|\mathbf{q}|^{-3}$.
It is found that the tail beyond $q_{c_{2}}$ contributes less than $1\%$ to the final value of the integral and can thus be neglected.

Finally, the integral over $|\mathbf{p}|$ is also divided in two intervals, namely, $[0,p_{c_{1}}]$ and $[p_{c_{1}},p_{c_{2}}]$, where $p_{c_{1}}=q_{c}$ while
$p_{c_{2}}=5k_{F}$ when $a_{F}<0$ and $p_{c_{2}}=7k_{F}$ when $a_{F}>0$.
Even in this case, the integrand turns out to have a minimum at about $p_{c_{1}}$, past which it decays to zero like $|\mathbf{p}|^{-4}$, while the tail beyond $p_{c_{2}}$ is found to contribute at most about $2$-$3\%$ to the final value of the integral.
                                                                                                                                                                                                                                                                                                                                                                                                      
\section{CONTRIBUTION OF THE POPOV AND GMB DIAGRAMS TO THE SCATTERING LENGTH OF COMPOSITE BOSONS IN THE BEC LIMIT}
\label{sec:appendix-B}

In Section~\ref{sec:G-MB-BCS-BEC}, the expressions of the Popov and GMB bosonic-like self-energies were obtained analytically in the BEC (strong-coupling) limit.
We now show that these expressions are related to the scattering length $a_{B}$ for the low-energy scattering between composite bosons that form out of fermion pairs in the BEC limit.
To this end, we rewrite the result (\ref{Popov-self-energy-approximate-BEC}) for the Popov bosonic-like self-energy in the BEC limit in the form:
\begin{equation}
\Sigma_{\mathrm{Popov}}^{\mathrm{B}} = - \frac{m k_{F}^{3} a_{F}^{2}}{6 \pi^{2}} = - \left( \frac{m^{2} a_{F}}{8 \pi} \right) \, \frac{8 \pi (2 a_{F})}{2m} \, \frac{n}{2} \, ,
\label{Popov-self-energy-approximate-BEC-rewritten}
\end{equation}
\noindent
as well as the result (\ref{GMB-approximation-BEC}) for the GMB bosonic-like self-energy in the BEC limit in the form:
\begin{equation}
\Sigma_{\mathrm{GMB}}^{\mathrm{B}} = \frac{16 \tilde{I}}{\pi^{2}} \frac{m k_{F}^{3} a_{F}^{2}}{6 \pi^{2}} = \left( \frac{m^{2} a_{F}}{8 \pi} \right) \frac{16 \tilde{I}}{\pi^{2}} \,\, \frac{8 \pi (2 a_{F})}{2m} \, \frac{n}{2} \, .
\label{GMB-approximation-BEC-rewritten}
\end{equation}
\noindent 
Here, $2m$ is the mass $m_{B}$ of a composite boson and $n/2$ the density $n_{B}$ of the system of composite bosons. 

In both expressions (\ref{Popov-self-energy-approximate-BEC-rewritten}) and (\ref{GMB-approximation-BEC-rewritten}), the factor $m^{2} a_{F}/(8 \pi)$ is required to comply with the structure of the pair propagator (\ref{pair-propagator-BEC-limit}), which in the (extreme) BEC limit acquires the polar form:
\begin{equation}
\Gamma_{0}(\mathbf{q},\Omega_{q}) = - \left( \frac{8 \pi}{m^{2} a_{F}} \right) \,\frac{1}{i \Omega_{q} - \xi^{B}_{\mathbf{q}}} \, .
\label{polar-pair-propagator-BEC-limit}
\end{equation}
\noindent
Apart from the factor $- 8 \pi/(m^{2} a_{F})$, the expression (\ref{polar-pair-propagator-BEC-limit}) has the structure of the single-particle propagator of free (point-like) bosons with dispersion relation $\xi^{B}_{\mathbf{q}} = \mathbf{q}^{2}/(4m) - \mu_{B}$.
In addition, when the interaction is taken into account in a low-density gas of (point-like) bosons, this free-boson propagator is complemented by a self-energy of the form
$8 \pi a_{B} n_{B}/m_{B}$ \cite{Popov-1987}.

\begin{figure}[t]
\begin{center}
\includegraphics[width=5.0cm,angle=0]{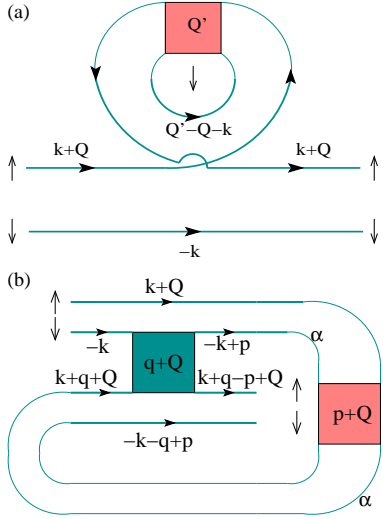}
\caption{(Color online) Alternative drawing of (a) the Popov diagram of Fig.~\ref{Figure-2}(a) and (b) the GMB diagram of Fig.~\ref{Figure-4}(a), which in both cases evidences a substructure 
                                    (in thick blue color) contributing to the scattering length $a_{B}$ for composite bosons when $Q=0$ that survives in the limit $n \rightarrow 0$.
                                    To better highlight the correspondence with $a_{B}$, the fermionic spins have been explicitly indicated.
                                    The pair propagator $\Gamma_{0}$ responsible for the presence of a single factor of the density $n$ in both Eqs.~(\ref{Popov-self-energy-approximate-BEC-rewritten})
                                    and (\ref{GMB-approximation-BEC-rewritten}) is also shown in pink color.}
\label{Figure-11}
\end{center} 
\end{figure}

Upon translating this information back in the language of composite bosons, we then conclude that to the Popov expression (\ref{Popov-self-energy-approximate-BEC-rewritten}) there corresponds the value $a_{B} = 2 a_{F}$ of the scattering length $a_{B}$ of composite bosons in terms of the scattering length $a_{F}$ of the constituent fermions.
This value of $a_{B}$ amounts to treating two-fermion scattering at the level of the Born approximation \cite{PS-2000}.
The GMB expression (\ref{GMB-approximation-BEC-rewritten}), on the other hand, contributes the value $- (16 \tilde{I}/\pi^{2}) 2 a_{F} \simeq - 0.842 a_{F}$ to the scattering length $a_{B}$,
with a different sign with respect to the Popov contribution.
In conclusion, when combined together the Popov and GMB contributions \emph{alone} yield the approximate value $a_{B} \simeq 1.158 a_{F}$, which has to be compared with the exact result $a_{B} \simeq 0.6 a_{F}$ obtained when \emph{all} possible scattering processes among composite bosons are taken into account \cite{Petrov-2005,Brodsky-2005}.

In this context, it is instructive to identify directly in the Popov diagram of Fig.~\ref{Figure-2}(a) and the GMB diagram of Fig.~\ref{Figure-4}(a) the sub-structures, which are associated 
with the processes that contribute to the scattering length $a_{B}$ of composite bosons.
This can be readily done by redrawing these diagrams in the alternative way shown in Fig.~\ref{Figure-11}, which makes these processes evident.
In addition, in Fig.~\ref{Figure-11}(b) we identify the two fermionic propagators $G_{0}$ (by denoting them with the label $\alpha$), which would give rise to the particle-hole bubble in the BCS limit discussed
in subsection \ref{sec:G-MB-BCS-BEC}-E, but which in the BEC limit here considered take part to the effective interaction between composite bosons.



\vspace{1.0cm}

\begin{thebibliography}{99}

\bibitem{BCS-1957} J. Bardeen, L. N. Cooper, and J. R. Schrieffer, \emph{Theory of Superconductivity}, Phys. Rev. {\bf 108}, 1175 (1957).

\bibitem{GMB-1961} L. P. Gorkov and T. M. Melik-Barkhudarov, \emph{Contribution to the theory of superfluidity in an imperfect Fermi gas}, Sov. Phys. JETP {\bf 13}, 1018 (1961)
                                  [Zh. Eksp. Teor. Fiz. {\bf 40}, 1452 (1961)].

\bibitem{Galitskii-1958} V. M. Galitskii, \emph{The energy spectrum of a non-ideal Fermi gas}, Sov. Phys. JETP {\bf 7}, 104 (1958) [Zh. Eksp. Teor. Fiz. {\bf 34}, 151 (1958)].

\bibitem{Regal-2003} C. A. Regal and D. S. Jin, \emph{Measurement of positive and negative scattering lengths in a Fermi gas of atoms}, Phys. Rev. Lett. {\bf 90}, 230404 (2003).

\bibitem{Schrieffer-1964} J. R. Schrieffer, \emph{Theory of Superconductivity} (Benjamin, New York, 1964).

\bibitem{Heiselberg-2000} H. Heiselberg, C. J. Pethick, H. Smith, and L. Viverit, \emph{Influence of induced interactions on the superfluid transition in dilute Fermi gases}, 
                                          Phys. Rev. Lett. {\bf 85}, 2418 (2000).

\bibitem{Combescot-1999} R. Combescot, \emph{Trapped $^{6}$Li: A high $T_{c}$ superfluid?}, Phys. Rev. Lett. {\bf 83}, 3766 (1999).

\bibitem{Petrov-2003} D. S. Petrov, M. A. Baranov, and G. V. Shlyapnikov, \emph{Superfluid transition in quasi two-dimensional Fermi gases}, Phys. Rev. A {\bf 67}, 031601(R) (2003).

\bibitem{Baranov-2008} M. A. Baranov, C. Lobo, and G. V. Shlyapnikov, \emph{Superfluid pairing between fermions with unequal masses}, Phys. Rev. A {\bf 78}, 033620 (2008).

\bibitem{Resende-2012} M. A. Resende, A. L. Mota, R. L. S. Farias, and H. Caldas, \emph{Finite temperature phase diagram of quasi-two-dimensional imbalanced Fermi gases beyond mean-field}, 
                                        Phys. Rev. A {\bf 86}, 033603 (2012).

\bibitem{Kim-2009} D.-H. Kim, P. T\"orma, and J.-P. Martikainen, \emph{Induced interactions for ultracold Fermi gases in optical lattices}, Phys. Rev. Lett. {\bf 102}, 245301 (2009).

\bibitem{Bulgac-2006} A. Bulgac, J. E. Drut, and P. Magierski, \emph{Spin 1/2 fermions in the unitary regime: A superfluid of a new type}, Phys. Rev. Lett. {\bf 96}, 090404 (2006).

\bibitem{Yu-2009} Z-Q. Yu, K. Huang, and L. Yin, \emph{Induced interaction in a Fermi gas with a BEC-BCS crossover}, Phys. Rev. A {\bf 79}, 053636 (2009).

\bibitem{Ruan-2013} X-X. Ruan, H. Gong, L. Du, W-M. Sun, and H-S. Zong, \emph{Effect of the induced interaction on the superfluid-transition temperature of ultracold Fermi gases within 
                                  the T-matrix approximation}, Phys. Rev. A {\bf 87}, 043608 (2013).

\bibitem{Lee-2017} J. Lee and D-H. Kim, \emph{Induced interactions in the BCS-BEC crossover of two-dimensional Fermi gases with Rashba spin-orbit coupling}, Phys. Rev. A {\bf 95}, 033609 (2017).

\bibitem{Floerchinger-2008} S. Floerchinger, M. Scherer, S. Diehl, and C. Wetterich, \emph{Particle-hole fluctuations in BCS-BEC crossover}, Phys. Rev. B {\bf 78}, 174528 (2008).

\bibitem{Pieri-2005} P. Pieri and G. C. Strinati, \emph{Popov approximation for composite bosons in the BCS-BEC crossover}, Phys. Rev. B {\bf 71}, 094520 (2005).

\bibitem{Popov-1987} V. N. Popov, \emph{Functional Integrals and Collective Excitations} (Cambridge University Press, Cambridge, England 1987), Chapt.~6.

\bibitem{Schulze-2001} H.-J. Schulze, A. Polls, and A. Ramos, \emph{Pairing with polarization effects in low-density neutron matter}, Phys. Rev. C {\bf 63}, 044310 (2001).

\bibitem{Cao-2006} L. G. Cao, U. Lombardo, and P. Schuck, \emph{Screening effects in superfluid nuclear and neutron matter within Brueckner theory}, Phys. Rev. C {\bf 74}, 064301 (2006).

\bibitem{Sa_de_Melo-1993} C. A. R. S\'{a} de Melo, M. Randeria, and J. R. Engelbrecht, \emph{Crossover from BCS to Bose superconductivity: Transition temperature and time-dependent 
                                             Ginzburg-Landau theory}, Phys. Rev. Lett. {\bf 71}, 3202 (1993).

\bibitem{NSR-1985} P. Nozi\`{e}res and S. Schmitt-Rink, \emph{Bose condensation in an attractive fermion gas: From weak to strong coupling superconductivity}, J. Low Temp. Phys. {\bf 59}, 195 (1985).

\bibitem{PPSC-2002} A. Perali, P. Pieri, G. C. Strinati, and C. Castellani, \emph{Pseudogap and spectral function from superconducting fluctuations to the bosonic limit}, Phys. Rev. B {\bf 66}, 024510 (2002).

\bibitem{Thouless-1960} D. J. Thouless, \emph{Perturbation theory in statistical mechanics and the theory of superconductivity}, Ann. Phys. {\bf 10} 553 (1960).

\bibitem{footnote-1} Here and in the following, we refer to ``bosonic-like self-energy''  $\Sigma^{\mathrm{B}}$ as those diagrams in the particle-particle channel which dress the 
                                ``bare'' pair propagator $\Gamma_{0}$. For this reason, a pre-factor $-m^{2}a_{F}/(8 \pi)$ has to be isolated from the expression of $\Sigma^{\mathrm{B}}$ 
                                to recover the corresponding bosonic expression, as shown explicitly in Appendix~\ref{sec:appendix-B}.
                               
\bibitem{AGD-1963} A. A. Abrikosov, L. P. Gorkov, and I. E. Dzyaloshinski, \emph{Methods of Quantum Field Theory in Statistical Physics} (Dover Publ., New York, 1963).

\bibitem{PS-2008} C. J. Pethick and H. Smith, \emph{Bose-Einstein Condensation in Dilute Gases} (Cambridge Univ. Press, Cambridge, 2008), Chapt.~16.

\bibitem{Mahan-2000} G. D. Mahan, \emph{Many-Particle Physics} (Kluwer, New York, 2000), Chapt. 5.

\bibitem{Feynman-1951} R. P. Feynman, \emph{An operator calculus having applications in Quantum Electrodynamics}, Phys. Rev. {\bf 84}, 108 (1951). 

\bibitem{PS-2000} P. Pieri and G. C. Strinati, \emph{Strong-coupling limit in the evolution from BCS superconductivity to Bose-Einstein condensation}, Phys. Rev. B {\bf 61}, 15370 (2000).

\bibitem{FW}  A. L. Fetter and J. D. Walecka, \emph{Quantum Theory of Many-Particle Systems} (McGraw-Hill, New York, 1971), Sect. 25.



\bibitem{Perali-2004} A. Perali, P. Pieri, L. Pisani, and G. C. Strinati, \emph{BCS-BEC crossover at finite temperature for superfluid trapped Fermi atoms}, Phys. Rev. Lett. {\bf 92}, 220404 (2004).

\bibitem{Haussmann-1994} R. Haussmann, \emph{Properties of a Fermi liquid at the superfluid transition in the crossover region between BCS superconductivity and Bose-Einstein condensation}, 
                                             Phys. Rev. B {\bf 49}, 12975 (1994).

\bibitem{Haussmann-2007} R. Haussmann, W. Rantner, S. Cerrito, and W. Zwerger, \emph{Thermodynamics of the BCS-BEC crossover}, Phys. Rev. A {\bf 75}, 023610 (2007).

\bibitem{Baym-1999} G. Baym, J-P. Blaizot, M. Holzmann, F. Lalo\"{e}, and D. Vautherin, \emph{The transition temperature of the dilute interacting Bose gas}, Phys. Rev. Lett. {\bf 83}, 1703 (1999).

\bibitem{Bulgac-2008} A. Bulgac, J. E. Drut, and P. Magierski, \emph{Quantum Monte Carlo simulations of the BCS-BEC crossover at finite temperature}, Phys. Rev. A {\bf 78}, 023625 (2008).

\bibitem{Burovski-2008} E. Burovski, E. Kozik, N. Prokof'ev, B. Svistunov, and M. Troyer, \emph{Critical temperature curve in BEC-BCS crossover}, Phys. Rev. Lett. {\bf 101}, 090402 (2008).

\bibitem{Nascimbene-2010} S. Nascimb\`{e}ne, N. Navon, K. J. Jiang, F. Chevy, and C. Salomon, \emph{Exploring the thermodynamics of a universal Fermi gas}, Nature {\bf 463}, 1057 (2010).

\bibitem{Ku-2012} M. J. H. Ku, A. T. Sommer, L. W. Cheuk, and M. W. Zwierlein, \emph{Revealing the superfluid lambda transition in the universal thermodynamics of a unitary Fermi gas}, 
                              Science {\bf 335}, 563 (2012).

\bibitem{Horikoshi-2010} M. Horikoshi, S. Nakajima, M. Ueda, and T. Mukaiyama, \emph{Measurement of universal thermodynamic functions for a unitary Fermi gas}, Science {\bf 327}, 442 (2010).

\bibitem{Goulko-2010} O. Goulko and M. Wingate, \emph{Thermodynamics of balanced and slightly spin-imbalanced Fermi gases at unitarity}, Phys. Rev. A {\bf 82}, 053621 (2010).

\bibitem{Pisani-2004} P. Pieri, L. Pisani, and G. C. Strinati, \emph{BCS-BEC crossover at finite temperature in the broken-symmetry phase}, Phys. Rev. {\bf 70}, 094508 (2004).

\bibitem{Perali-2011} A. Perali, F. Palestini, P. Pieri, G. C. Strinati, J. T. Stewart, J. P. Gaebler, T. E. Drake, and D. S. Jin, \emph{Evolution of the normal state of a strongly interacting 
                                   Fermi gas from a pseudogap phase to a molecular Bose gas}, Phys. Rev. Lett. {\bf 106}, 060402 (2011).

\bibitem{Palestini-2013} F. Palestini and G. C. Strinati, \emph{Systematic investigation of the effects of disorder at the lowest order throughout the BCS-BEC crossover}, Phys. Rev. B {\bf 88}, 174504 (2013).

\bibitem{Gaebler-2010} J. P. Gaebler, J. T. Stewart, T. E. Drake, D. S. Jin, A. Perali, P. Pieri, and G. C. Strinati, \emph{Observation of pseudogap behaviour in a strongly interacting 
                                      Fermi gas}, Nat. Phys. {\bf 6}, 569 (2010).

\bibitem{Carlson-2005} J. Carlson and S. Reddy, \emph{Asymmetric two-component fermion systems in strong coupling}, Phys. Rev. Lett. {\bf 95}, 060401 (2005).

\bibitem{Vale-2017} S. Hoinka, P. Dyke, M. G. Lingham, J. J. Kinnunen, G. M. Bruun, and C. J. Vale, \emph{Goldstone mode and pair-breaking excitations in atomic Fermi superfluids}, 
                                Nat. Phys. {\bf 13}, 943 (2017).

\bibitem{Carlson-2010} A. Gezerlis and J. Carlson, \emph{Low-density neutron matter}, Phys. Rev. C {\bf 81}, 025803 (2010).


\bibitem{Petrov-2005} D. S. Petrov, C. Salomon, and G. V. Shlyapnikov, \emph{Scattering properties of weakly bound dimers of fermionic atoms}, Phys. Rev. A {\bf 71}, 012708 (2005).

\bibitem{Brodsky-2005} I. V. Brodsky, A. V. Klaptsov, M. Y. Kagan, R. Combescot, and X. Leyronas, \emph{Bound states of three and four resonantly interacting particles},
                                      JETP Lett. {\bf 82}, 273 (2005).

\end{thebibliography}
\end{document}